\documentclass[longauth]{aaEC}
\usepackage{graphicx}
\usepackage{natbib}
\usepackage{scalerel}
\usepackage{comment}
\usepackage{bm}
\usepackage[usenames,dvipsnames]{color} 
\usepackage{placeins}

\usepackage[table]{xcolor}

\usepackage{txfonts}

\usepackage[pdfencoding=auto,psdextra]{hyperref}
\hypersetup{
    colorlinks=true,
    linkcolor=blue,
    filecolor=magenta,      
    urlcolor=blue,
    citecolor=blue
}
\makeatletter
\renewcommand*\aa@pageof{, page \thepage{} of \pageref*{LastPage}}
\makeatother

\usepackage[utf8]{inputenc}

\usepackage[switch, modulo]{lineno}
\linenumbers

\usepackage{euclid}
\begin{document}
\nolinenumbers

\title{Parametric strong lensing model of the galaxy cluster Abell 2390 from \Euclid and MUSE observations}

\author{\normalsize
  D.~Abriola$^{1}$\thanks{\email{davide.abriola@unimi.it}},
  M.~Lombardi$^{1,3}$, 
  C.~Grillo$^{1,2}$,
  P.~Bergamini$^{1,3}$,
  P.~Rosati$^{3,4}$,
  M.~Meneghetti$^{3}$,
  A.~Bolamperti$^{6,8,7}$,
  A.~Acebron$^{9,10,2}$,
  G.~Granata$^{5,4,1}$,
  G.~Angora$^{4,11}$,
  H.~Atek$^{13}$,
  J.M.~Diego$^{89}$,
  G.~Congedo$^{14}$,
  R.~Gavazzi$^{13,15}$,
  Y.~Kang$^{16}$,
  M.~Montes$^{17,18}$,
  T.T.~Thai$^{15,19}$
  }

\institute{$^{1}$ Dipartimento di Fisica, Università degli Studi di Milano, Via Celoria 16, I-20133 Milano, Italy\\
$^{2}$ INAF-IASF Milano, via A. Corti 12, I-20133 Milano, Italy\\
$^{3}$ INAF-OAS, Osservatorio di Astrofisica e Scienza dello Spazio di Bologna, via Gobetti 93/3, I-40129 Bologna, Italy \\
$^{4}$ Dipartimento di Fisica e Scienze della Terra, Università degli Studi di Ferrara, via Saragat 1, I-44122 Ferrara, Italy\\
$^{5}$ Institute of Cosmology and Gravitation, University of Portsmouth, Burnaby Rd, Portsmouth PO1 3FX, UK\\
$^{6}$ Dipartimento di Fisica e Astronomia, Università degli Studi di Padova, Vicolo dell'Osservatorio 3, I-35122 Padova, Italy\\
$^{7}$ INAF-Osservatorio Astronomico di Padova, Vicolo dell'Osservatorio 5, I-35122 Padova, Italy\\
$^{8}$ European Southern Observatory, Karl-Schwarzschild-Str. 2, D-85748 Garching bei München, Germany\\
$^{9}$ Instituto de Fisica de Cantabria (IFCA), CSIC - Universidad de Cantabria, Avda. los Castros, s/n, E-39005 Santander, Spain\\
$^{10}$ Departamento de Fisica Moderna, Universidad de Cantabria, Avda. de los Castros s/n, E-39005 Santander, Spain\\
$^{11}$ INAF–Osservatorio Astronomico di Capodimonte, Via Moiariello 16, I-80131 Napoli, Italy\\
$^{12}$ INAF-OAT Osservatorio Astronomico di Trieste, via G. B. Tiepolo 11, I-34131 Trieste, Italy\\
$^{13}$ Institut d’Astrophysique de Paris, UMR 7095, CNRS, and Sorbonne Université, 98 bis boulevard Arago, 75014 Paris, France\\
$^{14}$ Institute for Astronomy, University of Edinburgh, Royal Observatory, Blackford Hill, Edinburgh EH9 3HJ, UK\\
$^{15}$ Aix-Marseille Université, CNRS, CNES, LAM, Marseille, France\\
$^{16}$ Department of Astronomy, University of Geneva, ch. d’Ecogia 16, 1290 Versoix, Switzerland \\
$^{17}$ Instituto de Astrofísica de Canarias, Calle Vía Láctea s/n, 38204, San Cristóbal de La Laguna, Tenerife, Spain \\
$^{18}$ Departamento de Astrofísica, Universidad de La Laguna, 38206, La Laguna, Tenerife, Spain \\
$^{19}$ National Astronomical Observatory of Japan, 2-21-1 Osawa, Mitaka, Tokyo 181-8588, Japan
}

\abstract 
{We present a new high-precision parametric strong lensing total mass reconstruction of the \Euclid Early Release Observations (ERO) galaxy cluster Abell 2390 at redshift $z = 0.231$. We include in this analysis 35 multiple images from 13 background sources, of which 25 are spectroscopically confirmed thanks to observations from the Multi Unit Spectroscopic Explorer (MUSE), spanning a redshift range from $z$ = 0.535 to $z$ = 4.877. After fully re-analysing the MUSE spectroscopy, we combined it with archival spectroscopic catalogues, thus allowing us to select 65 secure cluster members. We further complemented this sample with 114 photometric member galaxies, identified within the \Euclid VIS and NISP imaging down to magnitude $\HE$ = 23. We also measured the stellar velocity dispersions for 22 cluster members in order to calibrate the Faber--Jackson relation and hence the scaling relations for the sub-halo mass components. We tested and compared 11 total mass parametrisations of the galaxy cluster with increasing complexity. To do so, we employed the new parametric strong lensing modelling code \texttt{Gravity.jl}. Our best-fit total mass parametrisation is characterised by a single large-scale halo, 179 sub-halo components, and an external shear term. The reference model yields a mean scatter between the model-predicted and observed positions of the multiple images of \ang{;;0.32}. We were able to quantify the systematics arising from our modelling choices by taking advantage of all the different explored total mass parametrisations. 
When comparing our results with those from other lensing studies, we noticed an overall agreement in the reconstructed cluster total mass profile in the outermost strong lensing regime. The discrepancy in the innermost region of the cluster (a few kiloparsecs from the brightest cluster galaxy, where few or no strong lensing features are observed) could possibly be ascribed to the different data and modelling choices.}

\titlerunning{Parametric strong lensing model of the galaxy cluster Abell 2390 from \Euclid and MUSE observations}
\authorrunning{Abriola et al.}
\maketitle

\section{Introduction}

Strong gravitational lensing (SL) in galaxy clusters is one of the most powerful probes for reconstructing the total mass distribution of their inner dense cores, where multiple images and giant arcs are formed. Once the baryonic mass components have been properly and independently mapped, it is possible to infer the dark matter mass distribution \citep[see, e.g. ][]{bonamigo2017, bonamigo2018, annunziatella2017, mahler18, granata2022, meneghetti2017, meneghetti2020, meneghetti2022, meneghetti2023, lagattuta2019, furtak23}. Moreover, strong lensing in galaxy clusters can lead to the discovery and study of high-redshift sources \citep[][]{atek2, atek1, borsani}, the study of the intrinsic properties of background lensed galaxies \citep[][]{cava18, magana18, mestric22, vanzella24}, and independent probing of the expansion and the geometry of the Universe by means of a variety of techniques, from the use measured time delays between multiple images \citep{acebron, bergamini2024, grillo2024} to the use of family ratios \citep{jullo2010, acebron2017, caminha2022, grillo2024}. To achieve these aims, accurate lensing modelling based on both high-quality imaging and spectroscopy is required, and this need has motivated several SL-dedicated surveys, including  the Cluster Lensing And Supernova survey with \textit{Hubble} \citep[CLASH;][]{clash}, the \textit{Hubble} Frontier Fields \citep[HFF;][]{lotz2017}, and more recently the Beyond Ultra-deep Frontier Fields And Legacy Observations \citep[BUFFALO;][]{buffalo} and the Ultradeep NIRSpec and NIRCam ObserVations before the Epoch of Reionization \citep[UNCOVER;][]{uncover}. The synergy between these projects is aimed at revolutionising the field by helping to develop high-resolution lens models of unprecedented quality.

It is in this context that \textit{Euclid}'s Early Release Observations \citep[ERO;][]{cuillandre} program Magnifying Lens \citep{atekero} sits. \textit{Euclid} \citep[][]{mellier} is an ongoing European Space Agency (ESA) mission aimed at probing cosmology through different methods, ranging from galaxy clustering to weak lensing. In addition to these scopes, \textit{Euclid} is expected to increase the number of SL systems by two orders of magnitude by observing approximately $170\,000$ galaxy-scale strong lenses \citep{collett2015, avecedo2024}, roughly $2000$ lensed quasars, and thousands of SL features in galaxy clusters distributed in the redshift range $z\in[0.2, 2.0]$ \citep{mellier, boldrin2012, boldrin2016}. These observations will help provide accurate total mass profiles of such systems from the kiloparsec to the megaparsec scale thanks to the ability of the telescope to combine strong and weak lensing measurements in galaxy clusters \citep{meneghetti2010, meneghetti2014, merten2015, umetsu2014, umetsu2020} and better understanding the assembly history of these objects. These findings will also further test the predictions of the standard cosmological Lambda cold dark matter ($\mathrm{\Lambda CDM}$) model and alternative dark matter models. The ERO program Magnifying Lens was developed to target two galaxy clusters, Abell 2390 and Abell 2764. In this paper, we focus on the analysis of the first cluster.

Abell 2390 (hereafter A2390, redshift $z = 0.231$) is one of the richest systems in the Abell galaxy cluster sample \citep{pello, leborgne}. It was first selected in a search for arcs on the basis of its bright X-ray emission \citep{fort, xray, allen}. This system is characterised by elongated arcs \citep{melli, feix, olm, richard2021}, and it has an estimated mass of $M_\mathrm{200} \simeq 1.06 \times 10^{15} \, \mathrm{M_\odot}$, obtained through a weak lensing analysis \citep{okabe}.\footnote{We note that $M_\mathrm{200}$ is the total mass enclosed within a sphere with a radius inside which the total mass density of the cluster is 200 times the critical density of the Universe, at the redshift of the cluster.} This cluster was previously subject to several SL studies based on ground and \textit{Hubble} Space Telescope (HST) imaging \citep{pierre96, pello99, swinbank06} and, thanks to novel \Euclid observations, to weak lensing analyses as well \citep[][Diego et al., in prep.]{shrabbackWL}. Thanks to the \textit{Euclid} imaging data from these studies, combined with fully re-analysed archival deep spectroscopy obtained with the Multi Unit Spectroscopic Explorer \citep[MUSE;][]{muse10, muse14} at the Very Large Telescope (VLT), we are able to develop and present the first high-precision parametric strong lensing model of this cluster based on \textit{Euclid} imaging.

The paper is organized as follows. In Sect. \ref{Data} we present the imaging and spectroscopic datasets of A2390 that we used to build our lens models. Section \ref{model} contains the details of the lens models we explored in our work, including the selection of the cluster members and the multiple images, and the total mass parametrisations we studied. The results of the best-fit reference model are then presented and discussed in Sect. \ref{results}.

Throughout this paper we use the AB magnitude system and adopt a flat $\mathrm{\Lambda CDM}$ cosmology with $\Omega_\mathrm{m} = 0.3$, $\Omega_\Lambda = 0.7$, and $H_0 = 70\,\kmsMpc$. In this cosmological model, $1''$ corresponds to $3.67 \,  \mathrm{kpc}$ at the cluster redshift, $z = 0.231$. The adopted galaxy cluster centre is $\mathrm{RA} = \ang{328.4034183;;}$, $\mathrm{Dec} = \ang{17.6954744;;}$ (J2000.0).

\section{Data}
\label{Data}

\subsection{Imaging data}
\label{photometry}

\begin{figure*}
    \centering
    \includegraphics[width=1.0\linewidth]{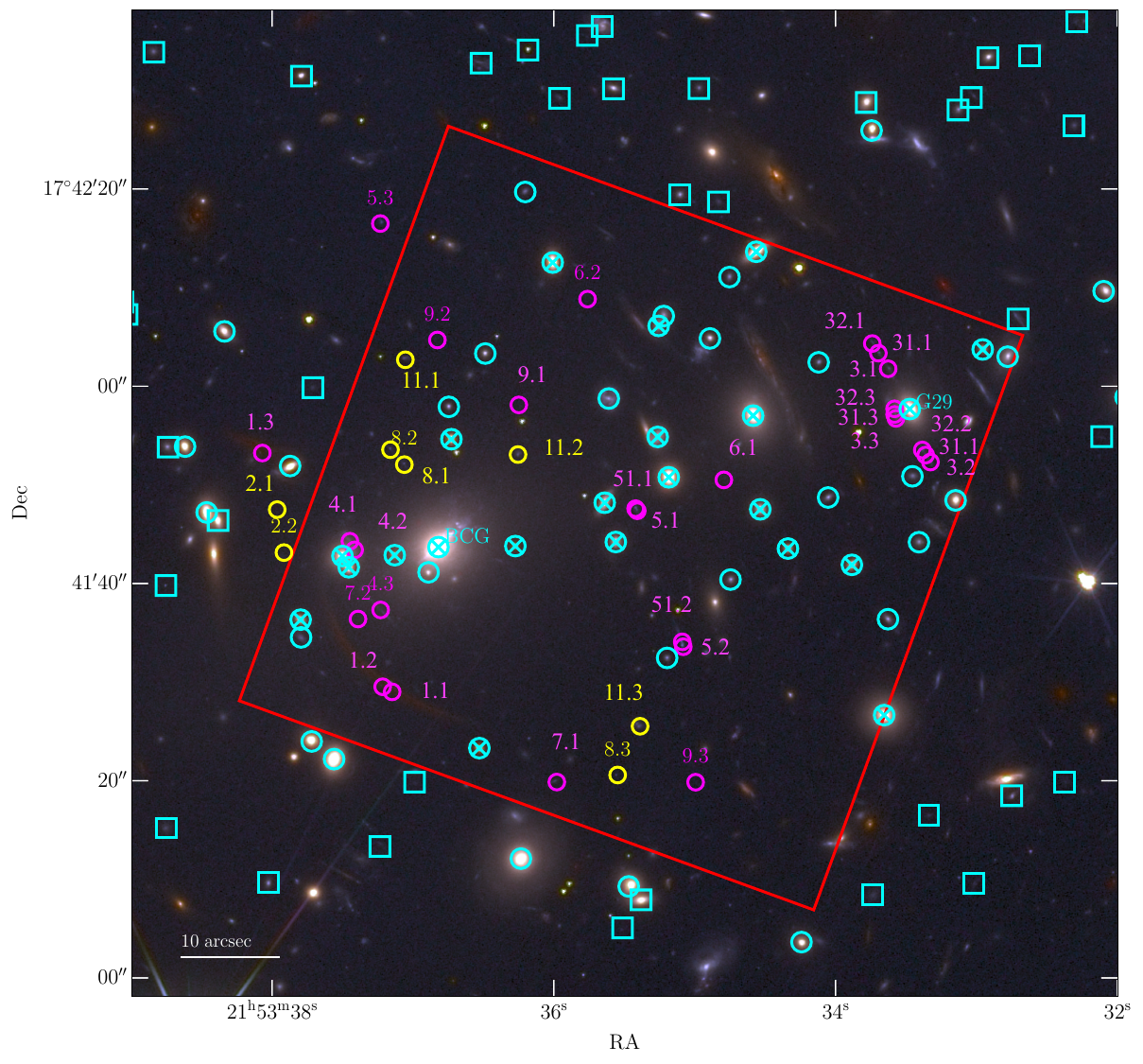}
    \caption{Colour-composite \Euclid image (red: $\HE$, green: $\JE+\HE$, blue: $\IE$) of the galaxy cluster A2390. The MUSE footprint is shown in red. The spectroscopic (photometric) cluster members  are represented with cyan circles (boxes). The 22 cluster members for which we measured the stellar velocity dispersion are further marked with cyan crosses. The spectroscopically confirmed (photometric) multiple images included in our analysis are also shown in magenta (yellow). The multiple images are also labelled with their ID (see Table \ref{tab:mi}). The positions of the BCG and galaxy G29 (see Sect. \ref{results}) are also marked.}
    \label{fig:image}
\end{figure*}

\noindent The lens cluster A2390 was targeted on 28 November 2023 with both the VISible camera \citep[VIS;][]{vis} and the Near-Infrared Spectrometer and Photometer \citep[NISP;][]{nisp} instruments on board the \textit{Euclid} satellite as part of the performance-verification phase of the mission. Three reference observing sequences \citep[ROS;][]{scaramella} were obtained for this galaxy cluster, resulting in three times the nominal \textit{Euclid} Wide Field Survey exposure time. Each ROS had a duration of 70.2 minutes, for a total exposure time of $3.30$ hours. The observations were carried out in all of \textit{Euclid}'s four filters ($\IE$ on VIS and $\JE$, $\YE$, and $\HE$ on NISP), covering an area of approximately $0.75 \, \mathrm{deg}^2$. The resolution of the VIS (NISP) image is $\ang{;;0.1}$ ($\ang{;;0.3}$, respectively), with $5\sigma$ point-like sources limiting magnitudes equal to 26.5 for the VIS and 24.5 for the NISP camera. The full data reduction pipeline is described in detail in \cite{cuillandre}. A photometric catalogue was extracted from the imaging with the public software \texttt{SourceXtractor++} \citep{se++1, se++2}. A more in-depth description of the photometric measurements is given in \cite{atekero}. The astrometric calibration was carried out with the software \texttt{SCAMP} \citep{scamp}. The reference catalogue used for the calibration of the ERO data was \textit{Gaia} Data Release 3 \citep{gaia}. Figure \ref{fig:image} shows an RGB colour-composite extract of A2390.

We also made use of archival multi-band HST imaging of this cluster (see Sect. \ref{spectroscopy}) obtained with the Wide Field and Planetary Camera 2 (WFPC2) within the CYCLE4 HIGH HST Program 5352 \citep[P.I.: Fort,][]{beze}. The lens cluster A2390 was observed by HST on 10 December 1994 for a total exposure time of $2\,100$ seconds.

\subsection{Spectroscopic data}
\label{spectroscopy}

\noindent We made use of archival VLT/MUSE \citep{Bacon2012} data of A2390, which was observed in the Wide Field Mode for a total integration time of 2 hours in September 2014 within the GTO Program 094.A-0115 \citep[P.I.: Richard,][]{richard2021}.
We employed the datacube available on the ESO Science portal, reduced by the Quality Control Group at the European Southern Observatory (ESO). The reduction process consisted in the removal of instrument signatures, sky subtraction, and combination of the two observing blocks as the latest step. Additionally, we enhanced the background sky subtraction by means of the Zurich Atmosphere Purge \citep[\texttt{ZAP};][]{Soto2016} tool. The MUSE footprint is shown in red in Fig. \ref{fig:image}. 
The resulting datacube spans the wavelength range from $4750 \, \AA$ to $9350 \, \AA$, with a constant sampling of $1.25 \, \mathrm{\AA/pix}$; covers a $1\arcmin \times 1\arcmin$ field of view (FoV); and has a spatial sampling of $\ang{;;0.2}\mathrm{/pix}$. It has a median point spread function full width half maximum of approximately $\ang{;;0.83}$. We registered the astrometry with respect to the HST/F814W image of the cluster.

We built a spectroscopic catalogue of the sources in the MUSE FoV following the procedure adopted in \citet{Caminha2019}. We considered a cutout of the HST/F814W image corresponding to the MUSE pointing imprint, and we ran {\tt SExtractor} \citep[{\tt v2.28.0},][]{Bertin1996} to detect all the included sources. We used the detected positions as the centres of circular apertures with a radius of \ang{;;0.8}, within which we extracted a spectrum. This aperture was chosen as to include the majority of the flux of the relative source while reducing the contamination of angularly close-by sources. To this catalogue we also added the sources that do not present a HST/F814W continuum but clear emission lines. These were identified through visual and automatic inspection using the Cube Analysis and Rendering Tool for Astronomy \citep[{\tt CARTA};][]{carta_2021}. We extracted their spectra within circular apertures with a radius of \ang{;;0.8} centred on the luminosity peak detected in narrow-line images obtained by collapsing the datacube around the wavelengths of the detected lines. 

We measured the redshift values for the objects in the catalogue making use of the {\tt Marz} \citep{Hinton2016Marz} software by identifying spectral features such as emission and absorption lines and continuum breaks through both automatic and visual analyses. We labelled each redshift measurement with a quality flag (QF) defined to be equal to one for tentative measurements, equal to two for possible measurements based on faint spectral features, equal to three for secure measurements based on multiple features, and equal to nine for measurements based on a single emission line. In some cases, if it was possible to characterise the single emission line (e.g. by observing a doublet or the typical asymmetric shape of the $\mathrm{Ly}\alpha$ line), we converted ${\rm QF}=9$ objects to ${\rm QF}=3$. 
The final MUSE catalogue contains 96 objects with ${\rm QF}>2$ redshift measurement divided into 6 stars or Galactic objects, 36 cluster members, and 54 background objects, including 25 multiple images from 10 background sources.

Since the MUSE pointing is only restricted to the core of the galaxy cluster (see Fig. \ref{fig:image}), we also considered ancillary archival spectroscopic measurements from several catalogues obtained with other instruments in order to complement the MUSE one. Specifically, we used the catalogues by \cite{sohn2020}, based on the SDSS Data Release 14 \citep{sdss} and the 6.5m Multiple Mirror Telescope (MMT) Hectospec spectrograph \citep{hectospec}, and the catalogue by \cite{abraham} based on the Multi Object Spectograph (MOS) and the Subarcsecond Imaging Spectrograph (SIS) at the 3.6m Canada-France-Hawaii Telescope (CFHT). We cross-matched these catalogues in order to produce a final spectroscopic sample. When a source was present in more than one catalogue, we gave priority (if possible) to the redshift measurements obtained through our reduction of the MUSE datacube. 

\begin{figure}
    \centering
    \includegraphics[width=1.0\linewidth]{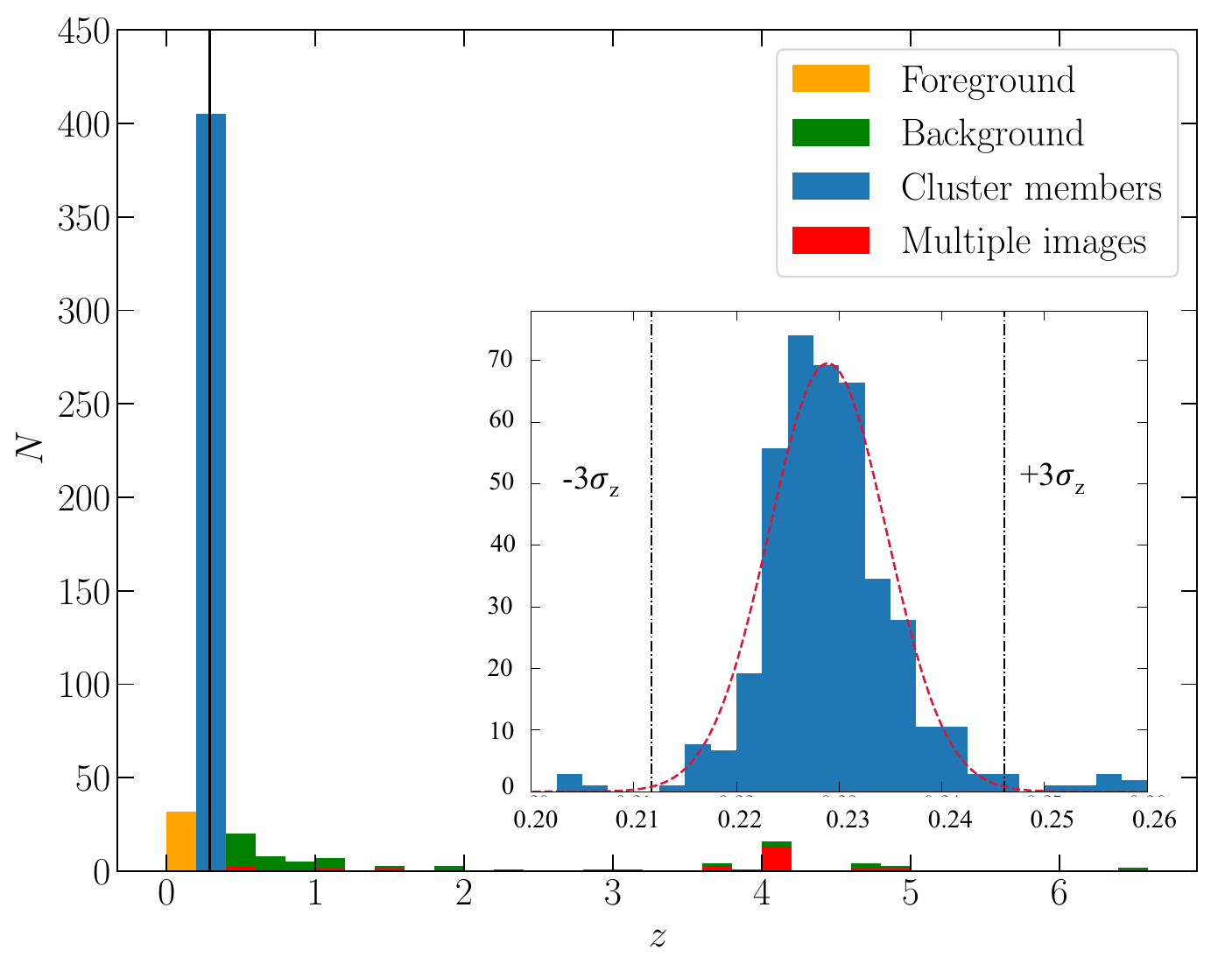}
    \caption{Spectroscopic redshift distribution of the objects in our final spectroscopic catalogue. Cluster members (i.e. lying in the redshift range from $z = 0.211$ to $z = 0.251$) are in blue, whereas foreground and background objects are in orange and green, respectively. Multiple images are depicted in red. The vertical black line locates the redshift of the galaxy cluster. The insert shows the cluster members selection illustrated in Sect. \ref{cluster-mems}. The dashed red line identifies the best-fit Gaussian distribution, whereas the vertical black dotted lines define an interval of $\pm 3\sigma_z$ around the median cluster redshift.}
    \label{fig:zdistribution}
\end{figure}

The final full spectroscopic catalogue consists of 592 sources within an effective FoV of approximately $40' \times 20'$. Of these, 36 are foreground galaxies or objects ($z \leq 0.211$), 405 are potential cluster members (i.e. lying in the redshift range $0.211 < z < 0.251$; see Sect. \ref{cluster-mems} for a detailed description of the selection of the cluster members), and 151 are background galaxies ($z \geq 0.251$), including 25 multiple images. Figure \ref{fig:zdistribution} displays the redshift distribution of these sources.

\section{Strong lensing models}
\label{model}
\noindent In this section we describe the SL total mass parametrisations of A2390 derived from the imaging and spectroscopic datasets presented in Sect. \ref{Data}. We performed our analysis with the newly developed software \texttt{Gravity.jl} \citep{gravity}, which allows for the reconstruction of the total mass distribution of a galaxy cluster by means of a Bayesian approach. This approach also enabled us to robustly compare the different mass parametrisations explored. To sample the posterior distribution of the models explored in our work, we employed the non-reversible parallel tempering algorithm \citep{pt, variational, piccioni} implementation included in \texttt{Gravity.jl}. We carried out the optimisations of the lens models in the simplified image-plane configuration. (For further details, see \cite{gravity}.)

In this work, we explored 11 total mass models characterised by the same set of cluster members and multiple images but a different total mass parametrisation. We labelled these total mass models as M1 to M11. A more detailed description of the different mass models is presented in Appendix \ref{appA}. In order to quantify the goodness of our lens models, we adopted two main indicators. First, we estimated the root mean square (RMS) separation between the observed and model-predicted positions of the multiple images. The RMS is defined as
\begin{equation}
    \Delta_\mathrm{RMS} = \sqrt{\frac{1}{N_\mathrm{im}} \, \sum_{i = 1}^{N_\mathrm{im}} \big|\mathbf{\Delta}_i \big|^2} = \sqrt{\frac{1}{N_\mathrm{im}} \, \sum_{i = 1}^{N_\mathrm{im}} \big|\vec{x}^{\mathrm{obs}}_i - \vec{x}^{\mathrm{pred}}_i\big|^2} \, ,
\end{equation}
where $\mathbf{\Delta}_i$ is the difference between the observed ($\vec{x}^{\mathrm{obs}}_i$) and predicted ($\vec{x}^{\mathrm{pred}}_i$) position of the $i$-th image, and $N_\mathrm{im}$ is the total number of multiple images considered in the model. 

Secondly, we estimated the evidence of the Bayes' theorem. This theorem is used to compute the probability of a set of parameters, $\bm{p}$, of a model, $M$, given the observed data, $\bm{d}$. It can be written as
\begin{equation}
    P(\bm{p}|\bm{d}, M) = \frac{\mathcal{L}(\bm{d}|\bm{p}, M) P(\bm{p}|M)}{P(\bm{d}|M)} \, ,
\end{equation}
where $\mathcal{L}(\bm{d}|\bm{p}, M)$ is the likelihood of the data given the parameters, $P(\bm{p}|M)$ is the prior (i.e. the a
priori knowledge of the parameters), and $P(\bm{p}|\bm{d}, M)$ is the posterior. The term $P(\bm{d})$ is a normalisation factor that does not depend on $\bm{p}$, called evidence. It plays a crucial role in quantifying the goodness of a model in the Bayesian framework. Indeed, the probability, $P(M|\bm{d})$, that the model, $M$, is correct when given the data can be estimated by using Bayes’s theorem again. Thus we can write
\begin{equation}
    P(M|\bm{d}) = \frac{P(\bm{d}|M) P(M)}{P(\bm{d})} \, .
\end{equation}
Here, $P(M)$ is a prior over $M$, that is, our belief that $M$ is the correct model before accessing the data. To compute the normalisation term, $P(\bm{d})$, one would need to marginalise over all possible models, which is generally impossible. Nevertheless, we can estimate the ratio $P(M|\bm{d})/P(M'|\bm{d})$, where $M'$ is an alternative model,
\begin{equation}
    \frac{P(M|\bm{d})}{P(M'|\bm{d})} = \frac{P(\bm{d}|M) P(M)}{P(\bm{d}|M') P(M')} \, ,
\end{equation}
which quantifies the preference of $M$ over $M'$. When comparing different modelling choices, we assumed $P(M) = P(M')$. In Table~\ref{table:lens_models} we quote the RMS and the natural logarithm of the evidence for the different total mass models.

\subsection{Cluster members}
\label{cluster-mems}
\begin{figure}
    \centering
    \includegraphics[width=1.0\linewidth]{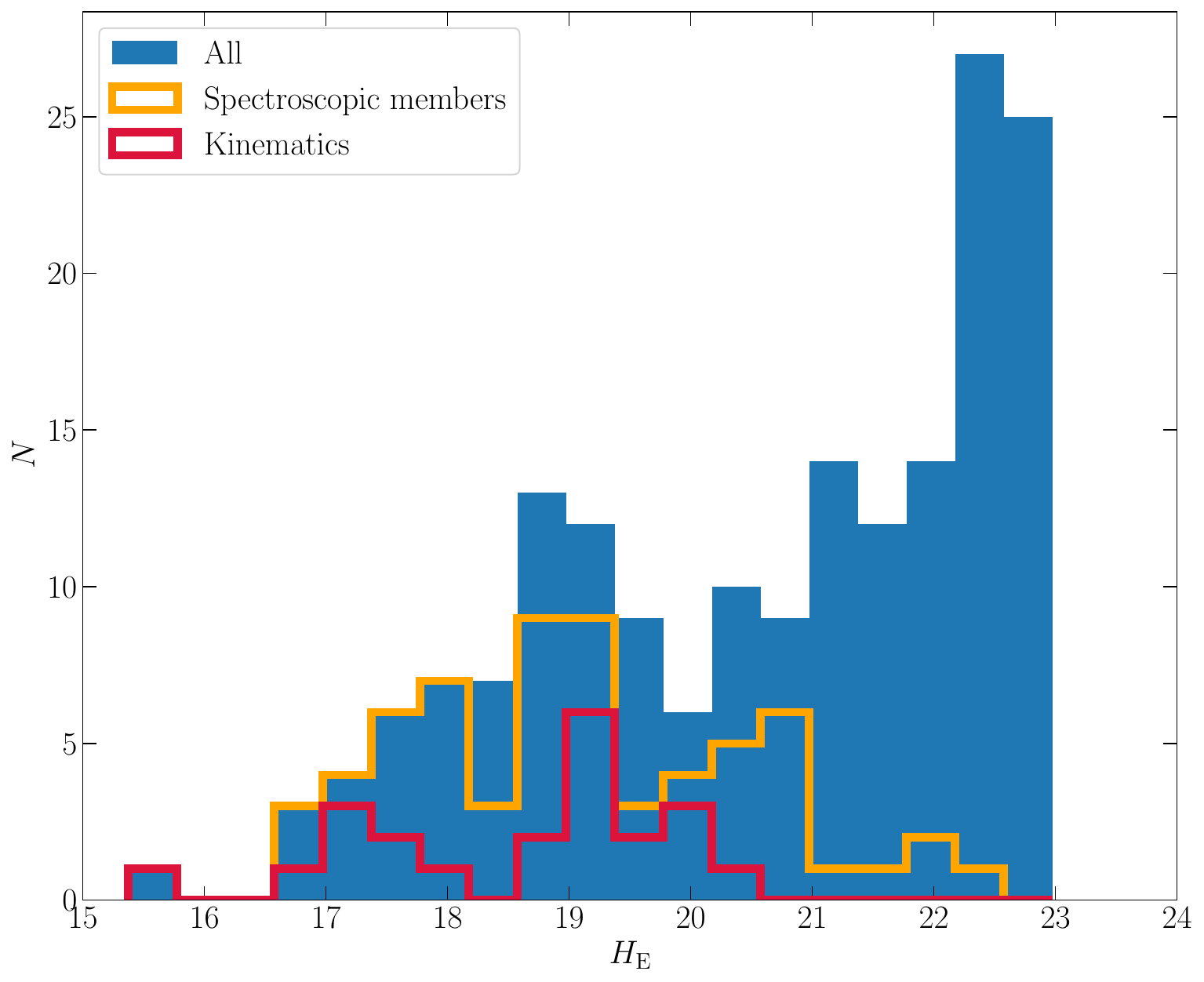}
    \caption{$\HE$ magnitude distribution of the 179 cluster members included in our models (blue). The 65 spectroscopically confirmed cluster members are highlighted in orange, whereas the sub-sample of the 22 galaxies for which we measured the central stellar velocity dispersion are in red.}
    \label{fig:mags}
\end{figure}

\noindent Cluster members were selected by exploiting both spectroscopic (see Sect. \ref{spectroscopy}) and multi-band \textit{Euclid} photometric (Sect. \ref{photometry}) information. We first identified a sample of spectroscopically confirmed cluster members starting from our spectroscopic catalogue as follows. We selected those galaxies lying within $\ang{;1.5;}$ from the centre of the cluster, which we assumed to be the brightest cluster galaxy (BCG). This choice was motivated by the fact that SL features (i.e. multiple images and giant arcs) are observed within 50 arcseconds from the BCG. The redshift distribution of these galaxies could then be fit with a Gaussian distribution, with mean and standard deviation values of $\hat z\pm\sigma_z = 0.231\pm0.005$, as depicted in the insert of Fig.~\ref{fig:zdistribution}. We thus identified 65 cluster members (cyan circles in Fig.~\ref{fig:image}), defined as those lying within $\pm 3 \, \sigma_z$ from $\hat z$ (this corresponds to about $\pm\,3700 \, \mathrm{km \, s^{-1}}$ around the cluster mean velocity). The magnitude distribution of these galaxies is represented with the orange solid line in Fig.~\ref{fig:mags}.

\begin{figure}
    \centering
    \includegraphics[width=1.0\linewidth]{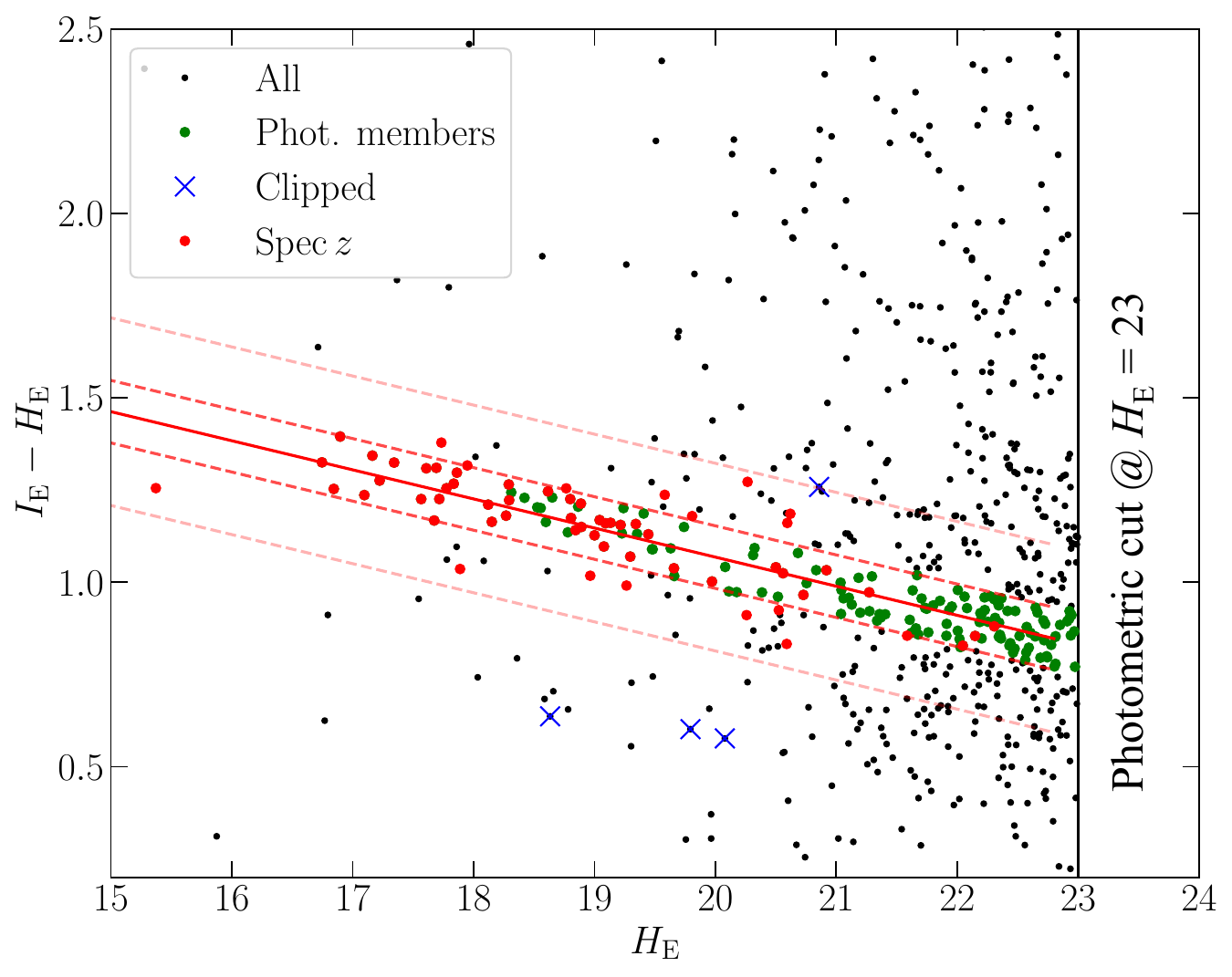}
    \caption{Colour-magnitude $\IE - \HE$ vs $\HE$ diagram. The red dots represent the 60 spectroscopically confirmed cluster members that we employed to fit the RCS, whereas the four cluster members we discarded due to the clipping are in blue. The green dots are the photometric galaxies added to our cluster member sample. The solid red line is the best-fit RCS, while the dotted ones define a range of $\pm \, \sigma_\mathrm{RCS}$ (used for the selection of the cluster members) and  $\pm \, 3\sigma_\mathrm{RCS}$ (used for the clipping) around the line, respectively.}
    \label{fig:rcs}
\end{figure}

We completed the spectroscopic sample by adding 114 further photometric (cyan boxes in Fig.~\ref{fig:image}), bright (\HE < 23) members by studying the distribution of the galaxies in a colour-magnitude $\IE - \HE$ versus $\HE$ diagram, as shown in Fig.~\ref{fig:rcs}. To evaluate the colours, we considered magnitudes measured within a radius of $\ang{;;0.5}$ and selected the galaxies lying within $\ang{;1.5;}$ from the BCG. We fitted the red cluster sequence (RCS) defined by the above-mentioned 65 spectroscopically confirmed cluster members  by means of a weighted linear regression. To do so, we used the Python package \texttt{ltsfit} \citep{capp}, which performs a least squares regression by iteratively clipping outliers \citep{clipp}. The result is presented in Fig.~\ref{fig:rcs}. We imposed a $3\sigma$ clipping, and thus discarded four members from the fit (represented by the blue crosses in Fig.~\ref{fig:rcs}; the red dots are the galaxies we employed for the fit). Given the best-fit, we extended our sample to include galaxies (the green dots in Fig.~\ref{fig:rcs}) within $\pm \, \sigma_\mathrm{RCS} = 0.1$ (the bold red dotted lines) from the RCS best-fit line (the solid red line), with $\sigma_\mathrm{RCS}$ as the intrinsic scatter of the data points around the best fit RCS. 

Following the procedure presented in \cite{granata23, granata24}, we measured the line-of-sight stellar velocity dispersion $\sigma_\mathrm{gal}$ for several cluster members from the MUSE data (their magnitude distribution is given by the red solid line in Fig.~\ref{fig:mags}). The velocity dispersion measurements were obtained using the spectral fitting code \texttt{penalized PiXel-Fitting} \citep[pPXF,][]{cappellari, cappellari23} to fit the observed spectra to a combination of 463 stellar spectral templates from the X-shooter Spectral Library (XSL) DR2 \citep{gonneau20}, convolved with a Gaussian line-of-sight velocity distribution. Following the methods of \citet{cappellari} and \cite{granata24}, we assessed the statistical uncertainty on the velocity dispersion values by generating $10\,000$ synthetic MUSE spectra and finding a relation between the statistical error on $\sigma_\mathrm{gal}$ and the signal-to-noise ratio $\mathrm{\langle S/N \rangle}$. We required a spectral $\mathrm{\langle S/N \rangle}$ greater than ten to ensure reliable velocity dispersion measurements, as shown by \citet{b19} and \cite{granata24}. The $\mathrm{\langle S/N \rangle}$ threshold required is an average for the whole spectrum, excluding the region masked for the potential influence of sky lines. The average refers to the $\mathrm{S/N}$ per spectral bin.

To measure the central stellar velocity dispersion of the cluster galaxies in their central regions, MUSE pixels were weighted by the surface brightness of the members in the \textit{Euclid} VIS band, degraded, and re-binned to match the MUSE point spread function. Spectra were extracted within $\ang{;;1.5}$ circular apertures centred on each galaxy, which significantly improved the spectral $\mathrm{\langle S/N \rangle}$. Light-weighting the cube resulted in values of the velocity dispersion approximately equivalent to those that would be measured within an aperture corresponding to the effective radius of each galaxy. Our final sample of cluster members with measured stellar velocity dispersion, presented in Table \ref{tab:vdispcat}, includes a total of 22 MUSE member galaxies. These are also marked with cyan crosses in Fig.~\ref{fig:image}. Three of the galaxies for which we measured $\sigma_\mathrm{gal}$ showed a spectrum potentially affected by light blending from nearby objects. Hence, for these galaxies we used the unweighted MUSE cube and extracted the spectrum from a fixed $\ang{;;0.6}$ aperture in order to reduce the contamination. These objects are marked in Table \ref{tab:vdispcat} with an asterisk. The $\HE$ magnitude distribution of the 179 cluster members included in our models is given in Fig.~\ref{fig:mags}. 

\subsection{Multiple image systems}
\noindent Our SL models comprise 35 multiple images from 13 background sources that were identified thanks to \Euclid and archival HST imaging. Of these, we confirmed 25 (corresponding to ten families) spectroscopically through our reduction of the MUSE datacube, spanning a redshift range between $z = 0.535$ and $z = 4.877$. These are represented with magenta circles in Fig.~\ref{fig:image}. Systems 1 and 5 are both composed of three multiple images, of which only two had a spectroscopic confirmation. Two families (8 and 11), despite being inside the MUSE footprint, do not show a prominent secure feature that allows us to determine their redshift. An additional family of multiple images (family 2) lies outside the footprint. The redshifts of these three  multiple-image systems (families 2, 8, and 11) were then optimized in our lens models with uninformative flat priors between $z = 0.24$ and $z = 10$. These photometric multiple images are shown with yellow circles in Fig.~\ref{fig:image}. The specifics of the multiple images included in our models are given in Table \ref{tab:mi}. For each image we report its coordinates, (if possible) its redshift as estimated through the reduction of the MUSE datacube, and the corresponding MUSE quality flag (we follow the same assignment legend introduced previously). In our work, the positions of the multiple images represent the observables for our lens models.

\begin{table}[!ht]
\caption{Coordinates and spectroscopic redshifts, with the corresponding
MUSE quality flag, of the multiple image systems used to build our models.}
\label{tab:mi}
\centering
\begin{tabular}{ccccc} 

  ID & $z_\mathrm{spec}$ & QF & RA & Dec\\ 
&  & & [$\mathrm{deg}$] & [$\mathrm{deg}$] \\
\noalign{\vskip 2pt}
 \hline
 \noalign{\vskip 2pt}
1.1	& 1.038 & 3  & 328.404788	& 17.691397\\
1.2	& 1.038 & 3 & 328.405068	& 17.691541\\
1.3	& 1.038 & $-$ & 328.408628 & 17.698124\\
2.1 & $-$ & $-$ & 328.408188 & 17.696531\\
2.2 & $-$ & $-$ & 328.407988 & 17.695320\\
3.1	& 4.048	& 3	& 328.390118	& 17.700497\\
3.2	& 4.048	& 3 & 328.388868	& 17.697870\\
3.3	& 4.048	& 3	& 328.389928	& 17.699370\\
31.1 & 4.048 &	3 & 328.390408	& 17.700937\\
31.2 & 4.048 & 3 & 328.389008	& 17.698067\\
31.3 & 4.048 & 3 	& 328.389928	& 17.699210\\
32.1 & 4.048 & 3  & 328.390588	& 17.701209\\
32.2 & 4.048 & 3  & 328.389108	& 17.698210\\
32.3 & 4.048 & 3  & 328.389888	& 17.699100\\
5.1	& 4.048	& 3	 & 328.397588	& 17.696557\\
5.2	& 4.048	& 3	 & 328.396178	& 17.692668\\
5.3	& 4.048	& $-$	 & 328.405138 & 17.704588\\
51.1 & 4.048 & 3  & 328.397538	& 17.696498\\
51.2 & 4.048 & 3	& 328.396208	& 17.692806\\
4.1	& 0.535 & 3	& 328.406038	& 17.695641\\
4.2	& 0.535 & 3	& 328.405898	& 17.695396\\
4.3	& 0.535 & 3	& 328.405235 & 17.693915\\
6.1	& 1.465 & 3	& 328.395096	& 17.697592\\
6.2	& 1.465 & 3	& 328.399096	& 17.702482\\
7.1	& 4.877 & 3	& 328.399918	& 17.688857\\
7.2	& 4.877 & 3	& 328.405798	& 17.693449\\
8.1 & $-$ & $-$ & 328.404838 & 17.698218\\
8.2 & $-$ & $-$ & 328.404428 & 17.697798\\
8.3 & $-$ & $-$ & 328.398120 & 17.689056\\
9.1	& 3.653	& 3	& 328.401036	& 17.699392\\
9.2	& 3.653	& 3	& 328.403273	& 17.701490\\
9.3	& 3.653	& 3	& 328.395725	& 17.688863\\
11.1 & $-$ & $-$ & 328.404398 & 17.700751\\
11.2 & $-$ & $-$ & 328.401066 & 17.698081\\
11.3 & $-$ & $-$ &328.397456 & 17.690428\\

\end{tabular}
\tablefoot{System 2 is outside the MUSE pointing, whereas families 8 and 11 did not display any secure feature, and hence their spectroscopic redshifts and quality flags are not assigned. Images 1.3 and 5.3 do not have spectroscopic confirmation, so we did not assign them a quality flag.
}
\end{table}

\subsection{Total mass parametrisation}
\noindent The total mass distribution of the galaxy cluster, or, equivalently, the total gravitational potential $\phi$ of the lens, is parametrised as the sum of three contributions: $N_\mathrm{h}$ large-scale smooth haloes representing the cluster dark matter component, $N_\mathrm{g}$ components for the cluster member galaxies, modelled as spherical dual pseudo-isothermal ellipsoid (dPIE) haloes, and a shear-like term which accounts for the presence of massive structures in the outskirts of the system and line-of-sight mass elements. Hence, the total cluster gravitational potential $\phi$ assumes the form
\begin{equation}
    \phi = \sum_{i=1}^{N_\mathrm{h}}\phi_i^{\mathrm{halo}} +  \sum_{j=1}^{N_\mathrm{g}}\phi_j^{\mathrm{gal}} +
    \phi^{\mathrm{shear}} \, .
\end{equation}

\subsection{Dark matter mass distribution}
\noindent In our work, the large-scale smooth haloes, representing mainly the dark matter component, are modelled as non-singular isothermal ellipsoids (NIEs). This mass distribution is characterised by six free parameters: the position on the sky $(x,y)$; the axis ratio ($q$), defined as the ratio between the semi-minor and semi-major axes of the projected ellipse; the position angle ($\theta$, computed clockwise from the north axis); the central velocity dispersion ($\sigma$); and the core radius, $r_\mathrm{core}$. The NIE is parametrised in \texttt{Gravity.jl} using the following parametrisation \cite[see e.g.][]{keeton_models}:
\begin{equation}
    \Sigma(R') = \frac{\sigma_\mathrm{v}^2}{2 G \sqrt{R'^2 + r_\mathrm{core}^2}} \, ,
\end{equation}
with
\begin{equation}
    R'^2 = x^2 + y^2/q^2 \, .
\end{equation}
In our analyses, we explored the total mass parametrisations including one (models M1--M4 and M9--M10), two (models M5--M8), or three (model M11) dark matter haloes.

\subsection{External shear term}
The external shear is described in the polar coordinates $(r, \theta)$ on the lens plane using the standard parametrisation \cite[see e.g.][]{keeton_models} as
\begin{equation}
    \phi^\mathrm{shear} (r, \theta) = \frac{|\gamma|}{2}r^2 \cos[2(\theta - \theta_\gamma)] \, .
\end{equation}
Here, $|\gamma|$ is the modulus of the shear and $\theta_\gamma$ is the position angle on the lens plane (also computed clockwise from the north axis), which yields the direction of the shear perturbation.

\subsection{Galaxy-scale mass distribution}
\begin{figure}
    \centering
    \includegraphics[width=1.0\linewidth]{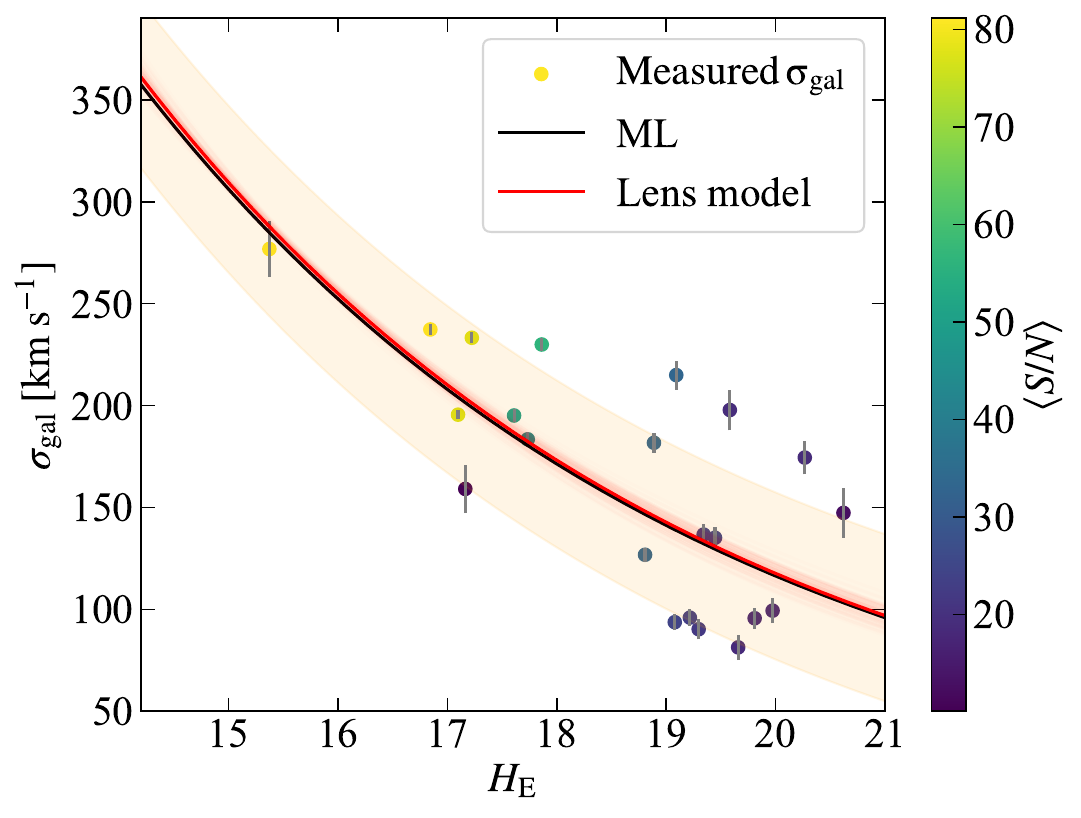}
    \caption{Measured stellar velocity dispersions of 22 MUSE spectroscopically confirmed cluster members as a function of their \Euclid $\HE$ magnitudes. Their colours encode the mean signal-to-noise ratio of galaxy spectra. The black solid line is the best-fit (maximum likelihood) of the scaling relation in Eq.~\eqref{sc1}. The light orange band corresponds to the best-fit mean scatter, $\Delta \sigma_\mathrm{ap}$, around the best-fit relation. The red solid curve corresponds to the relation in Eq.~\eqref{sc1} as obtained with the best-fit parameters of our reference model (see Sect. \ref{cluster-mems}). The light red area is estimated from 300 random values of $\sigma_\mathrm{ref}$ extracted from the Bayesian MCMC realisations of the reference model of this work.}
    \label{fig:fjfit}
\end{figure}

\begin{figure}
    \centering
    \includegraphics[width=1.0\linewidth]{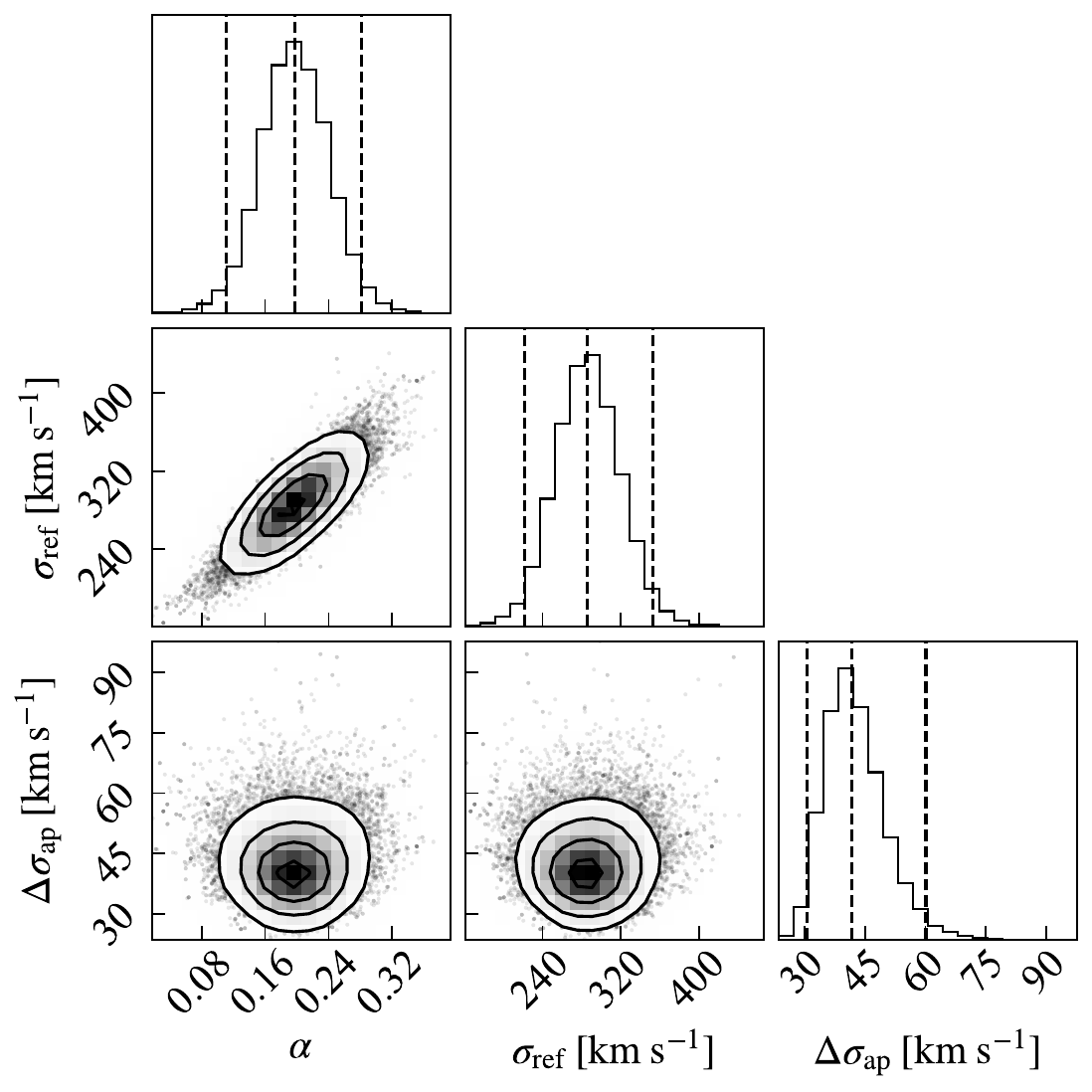}

    \caption{Posterior probability distribution for the Faber--Jackson relation calibration using the measurements of the velocity dispersion of 22 spectroscopically confirmed cluster members. The 16th, 50th, and 84th percentiles of the marginalised distributions for the slope ($\alpha$), normalisation ($\sigma_\mathrm{ref}$), and scatter around the scaling relation ($\Delta \sigma_\mathrm{ap}$) are displayed and shown as vertical dashed lines.}
    \label{fig:fjcorner}
\end{figure}

\noindent As mentioned previously, the cluster member galaxies (the sub-halo components of the galaxy cluster) are modelled in terms of spherical dPIE profiles. The spherical dPIE total mass distribution \citep{limousin, eliasdottir, b19} used to model them is described by three free parameters: the central velocity dispersion, $\sigma$; the core, $r_\mathrm{core}$; and the truncation radius, $r_\mathrm{cut}$. For the $i$-th cluster member, the central velocity dispersion, $\sigma_{\mathrm{gal}, i}$, and the truncation radius, $r_{\mathrm{cut},i}$, scale with its luminosity, $L_i$, according to the following relations \citep{jorge, nata, jullo07} -- which are introduced to decrease the number of free parameters of the lens model:
\begin{equation}
    \sigma_{\mathrm{gal}, i} = \sigma_{\mathrm{ref}}\left( \frac{L_i}{L_0}\right)^\alpha 
\label{sc1}
\end{equation}
and
\begin{equation}
    r_{\mathrm{cut},i} = r_{\mathrm{cut},\mathrm{ref}}\left( \frac{L_i}{L_0}\right)^\beta \, .
\label{sc2}
\end{equation}
Here, $L_0$ is a reference luminosity, which we assumed to be that of the BCG. We adopted the total $\HE$ magnitudes in the above scaling relations since they have been proved to be good proxies of the total mass of the galaxies \citep{grillo, b19}. The total magnitude of the BCG in the $\HE$ band is $\HE^\mathrm{ref} = 15.37$. Following the prescription by \citet{b19, b21}, we fitted the values of the slope, $\alpha$, and the normalisation, $\sigma_\mathrm{ref}$, of Eq.~\eqref{sc1}, which generalises the Faber--Jackson relation \citep{faber}, by employing the measured 22 stellar velocity dispersion velocities, $\sigma_\mathrm{gal}$ (see above). We also included an additional free parameter: the intrinsic scatter, $\Delta \sigma_\mathrm{ap}$, of the measured velocities around the scaling relation. We relied on the Bayesian approach shown in \cite{b19} and considered 12 walkers performing $10\,000$ steps each. We adopted the following uniform priors: $[\alpha_\mathrm{min}, \alpha_\mathrm{max}] = [0.0, 1.0]$, $[\sigma_\mathrm{ref, min}, \sigma_\mathrm{ref,max}] = [100\,\kms, 600\,\kms]$, and $[\Delta \sigma_\mathrm{ap, min}, \Delta \sigma_\mathrm{ap, max}] = [0\,\kms, 100\,\kms]$. The result is shown in Fig.~\ref{fig:fjfit}. The black solid line is the best-fit scaling relation (obtained with the set of values maximising the likelihood), whereas the light orange band represents the best-fit mean scatter, $\Delta \sigma_\mathrm{ap}$, around the relation. In Fig.~\ref{fig:fjcorner}, we show the marginalised posterior probability distribution for the Faber--Jackson relation calibration. With respect to previous works by \cite{b19, b21}, where the same approach was followed for different lens galaxy clusters, we found a lower median value of $\alpha$. We recovered $\alpha = 0.21 \pm 0.04$, whereas \cite{b19, b21} found values of $\alpha \in [0.27,0.30]$. 

To determine the value of $\beta$, we assumed a fixed scaling relation between the total mass, $M_{\mathrm{tot},i}$, of the $i$-th cluster member and its luminosity, namely, $M_{\mathrm{tot},i}/L_i \propto L_i^\gamma$. Under this assumption, we obtained
\begin{equation}
    \beta = \gamma - 2 \alpha + 1 \, ,
\end{equation}
with $\gamma = 0.2$ in order to be consistent with the observed fundamental plane relation.

Each galaxy is anchored at its measured position (see Tables \ref{tab:specmems1} and \ref{tab:photmems1}). Moreover, for each sub-halo (galactic) component, the value of the core radius was kept fixed to $\ang{;;0.005}$. Given the results of the fit of the generalised Faber--Jackson relation described above, in our models we adopted for $\sigma_{\mathrm{ref}}$ a Gaussian prior with a mean of $285\,\kms$ and a standard deviation of $41\,\kms$, whereas an uninformative prior was assumed on $r_{\mathrm{cut},\mathrm{ref}}$. In models M3--M4 and M7--M8, the BCG is excluded from the scaling relations and described independently in terms of a spherical dPIE profile, with both the velocity dispersion and the truncation radius as free parameters. In a similar manner, in models M9 and M10, galaxy G29 (see Fig.~\ref{fig:image}) is excluded from the scaling relations and modelled either with a spherical (M9) or an elliptical (M10) dPIE mass density profile. This modelling choice was made upon the motivations described in Sect.~\ref{lens_models_discussion}.

\begin{table}[h!]  
\centering
    \caption{Input and optimized parameter values of the reference lens model (M9) for the galaxy cluster A2390.}   
\scalebox{0.8}{
	\begin{tabular}{l c c c}

	\textbf{Mass component} & \textbf{Free parameter} & \textbf{Prior} & \textbf{Posterior}\\
    \noalign{\vskip 2.5pt}
	   \hline
       \noalign{\vskip 2pt}
	\textbf{Cluster halo} & $x_\mathrm{DM}\;[\mathrm{arcsec}]$ & $0\pm100$ & $2.76\,\pm\,0.08$ \\
    \noalign{\vskip 1pt}
      & $y_\mathrm{DM}\;[\mathrm{arcsec}]$ & $0\pm100$ & $1.20^{+0.06}_{-0.05}$\phantom{0} \\
      \noalign{\vskip 1pt}
      & $q_\mathrm{DM}$\phantom{0000000} & $0.25\div1.0$\phantom{0}\phantom{0}\phantom{0} & $0.73\,\pm\,0.01$ \\
      & $\theta_\mathrm{DM}\;[\mathrm{rad}]$\phantom{00} & $0\div\pi$\phantom{0}\phantom{0} & $1.1\,\pm\,0.1$ \\
      & \phantom{0}$\sigma_\mathrm{DM}\;[\kms]\phantom{0}$ & $500\div1500$ & $1195\,\pm\,12$\phantom{00} \\
      & $r_\mathrm{core,DM}\;[\mathrm{arcsec}]\phantom{00}$ & $0.01\div30.0$\phantom{0}\phantom{0} & $16.7\,\pm\,0.4$\phantom{0} \\
      \noalign{\vskip 2.5pt}
        \hline
        \noalign{\vskip 2.5pt}
        \textbf{G29} & \phantom{0}$\sigma_\mathrm{G29}\;[\kms]$\phantom{0} & $50\div500$\phantom{0} & $233^{+5}_{-3}$\phantom{00}\\
        \noalign{\vskip 2pt}
        & $r_\mathrm{cut,G29}\;[\mathrm{arcsec}]$\phantom{00} & $0.05\div25.0$\phantom{0}\phantom{0} & $18.2^{+4.3}_{-4.1}$\phantom{00} \\
        \noalign{\vskip 2.5pt}
       \hline
       \noalign{\vskip 2.5pt}
        \textbf{External shear} & $|\gamma|$\phantom{0000} & $0.0\div0.3$\phantom{0}\phantom{0} & $0.17\pm0.01$\\
        & $\theta_\gamma\;[\mathrm{rad}]$\phantom{0} & $0\div\pi$\phantom{0}\phantom{0} & $2.22\,\pm\,0.01$ \\
        \noalign{\vskip 2.5pt}
        \hline
        \noalign{\vskip 2.5pt}
        \textbf{Scaling relations} & \phantom{0}$\sigma_\mathrm{ref}\;[\kms]$\phantom{0} & $285\pm41$\phantom{0}\phantom{0}\phantom{0} & $288\pm5$\phantom{00}\\
        & $r_\mathrm{cut,ref}\;[\mathrm{arcsec}]\phantom{00}$ & $0.05\div30.0$\phantom{0}\phantom{0} & $16.1^{+1.3}_{-0.9}$\phantom{00} \\
        \noalign{\vskip 2.5pt}
        
	\end{tabular}}
    \tablefoot{The first column reports the mass component. The second column contains the parameters of the density profile used to describe the corresponding mass component. The third column shows the prior distributions adopted. When a flat prior on a free parameter value is considered, the boundaries of the prior separated by the $\div$ symbol are reported. In case of a Gaussian prior, the notation $a \pm b$ is adopted, with $a$ as the mean and $b$ as the standard deviation of the distribution, respectively. The $x$ and $y$ coordinates are measured with respect to the position of the BCG $(\mathrm{RA} = \ang{328.403408;;}$, $\mathrm{Dec} = \ang{17.695475;;})$. In the last column, we quote the median value and the 16th and 84th percentiles of the marginalised posterior distribution.}
	\smallskip
	\label{table:inout_lensing} 

\end{table}

\section{Results and discussion}
\label{results}

\subsection{Lens models}
\label{lens_models_discussion}
\noindent In Table \ref{table:lens_models} we briefly summarise the main properties of the 11 lens models explored in this work, M1 to M11, as well as the principal figures of merit adopted to quantify their goodness. We first explored models comprising one extended or cluster-scale dark matter halo (M1 to M4), and then we added a second halo (models M5 to M8). We also studied the impact of the inclusion (models M2, M4, M6, and M8) or exclusion (models M1, M3, M5, and M7) of an external shear term as well as the removal of the BCG from the scaling relations (models M3, M4, M7, and M8). As shown in Table \ref{table:lens_models}, the inclusion of an external shear term helps increase the accuracy of the models significantly by reducing the mean scatter between the observed and model-predicted positions of the multiple images by about $\ang{;;0.1}$. The exclusion of the BCG from the scaling relations does not seem to critically affect the figures of merit adopted in this work. Despite being the model with the lowest $\Delta_\mathrm{RMS}$ and highest value of the evidence, M6 predicts the second extended dark matter halo as lying approximately $50''$ north-west of the BCG in a region without any concentration of galaxies. Moreover, this second halo is predicted to lie in projection close to galaxy G29 (MUSE ID 29, see Fig. \ref{fig:image} and Table \ref{tab:specmems1}), which is surrounded by the elongated arc where the families 3, 31, and 32 of multiple images are observed. Furthermore, model M6 predicts for the second extended dark matter halo a velocity dispersion value more consistent with that of a cluster member (roughly $300\,\kms$). This motivated us to model galaxy G29 independently from the other cluster members by excluding it from the scaling relations. We therefore explored two further models, M9 and M10, where galaxy G29 is described in terms of either a spherical (model M9) or an elliptical (model M10) dPIE mass density profile. We also tested a final model, M11, characterised by three cluster-scale dark matter haloes, which despite the values of RMS and log evidence, can be discarded as well. Indeed, similarly to model M6, the second dark matter halo is predicted to lie in the same region $50''$ north-west of the BCG, with a velocity dispersion value consistent with that of a member galaxy. The third one is about $21''$ north-west of the BCG, also in a region with no observed cluster members. According to the figures of merit adopted in the paper, M9 stands out as the best-fit model. Hence, in the following, we discuss the results obtained with M9, which is the most physically plausible model and will be considered as the reference model of this work. 

The marginalised posterior distributions of the free parameters included in model M9 are given in Fig. \ref{fig:cornerplotM9}. Among the degeneracies observed, we noticed the expected one between the reference velocity dispersion and truncation radius of the scaling relations for the cluster members, namely, $M_{\mathrm{tot},i} \propto \sigma_{\mathrm{gal},i}^2 r_{\mathrm{cut},i}$. In addition, we noticed a correlation between the axis-ratio, $q_\mathrm{DM}$, and the shear, $\gamma$: As $q_\mathrm{DM}$ increases, which implies a rounder cluster-scale halo, the shear increases as well to compensate for the ellipticity.
The median value of $\sigma_\mathrm{ref}$, quoted in Table \ref{table:inout_lensing}, is consistent with the value recovered by calibrating the Faber-Jackson relation with the kinematics data. In Fig. \ref{fig:fjfit}, the red solid curve is the relation obtained by considering the results of M9. We note the difference between the shaded areas: The light orange band represents the intrinsic scatter, $\Delta \sigma_\mathrm{ap}$, around the relation, whereas the red line was obtained by randomly extracting 300 values of $\sigma_\mathrm{ref}$ from the Markov chain Monte Carlo (MCMC) realisations for M9. Additionally, the best-fit value of the velocity dispersion of G29 is in very good agreement with our measurement (see Table \ref{tab:vdispcat}). Interestingly, the values of the redshifts for families 8 and 11 are consistent with their tentative MUSE spectroscopic measurements. As far the external shear is concerned, its position angle is perpendicular to the direction of the NIE describing the cluster-scale halo.

\begin{figure*}[ht!]
    \centering
    \includegraphics[width=1.0\linewidth]{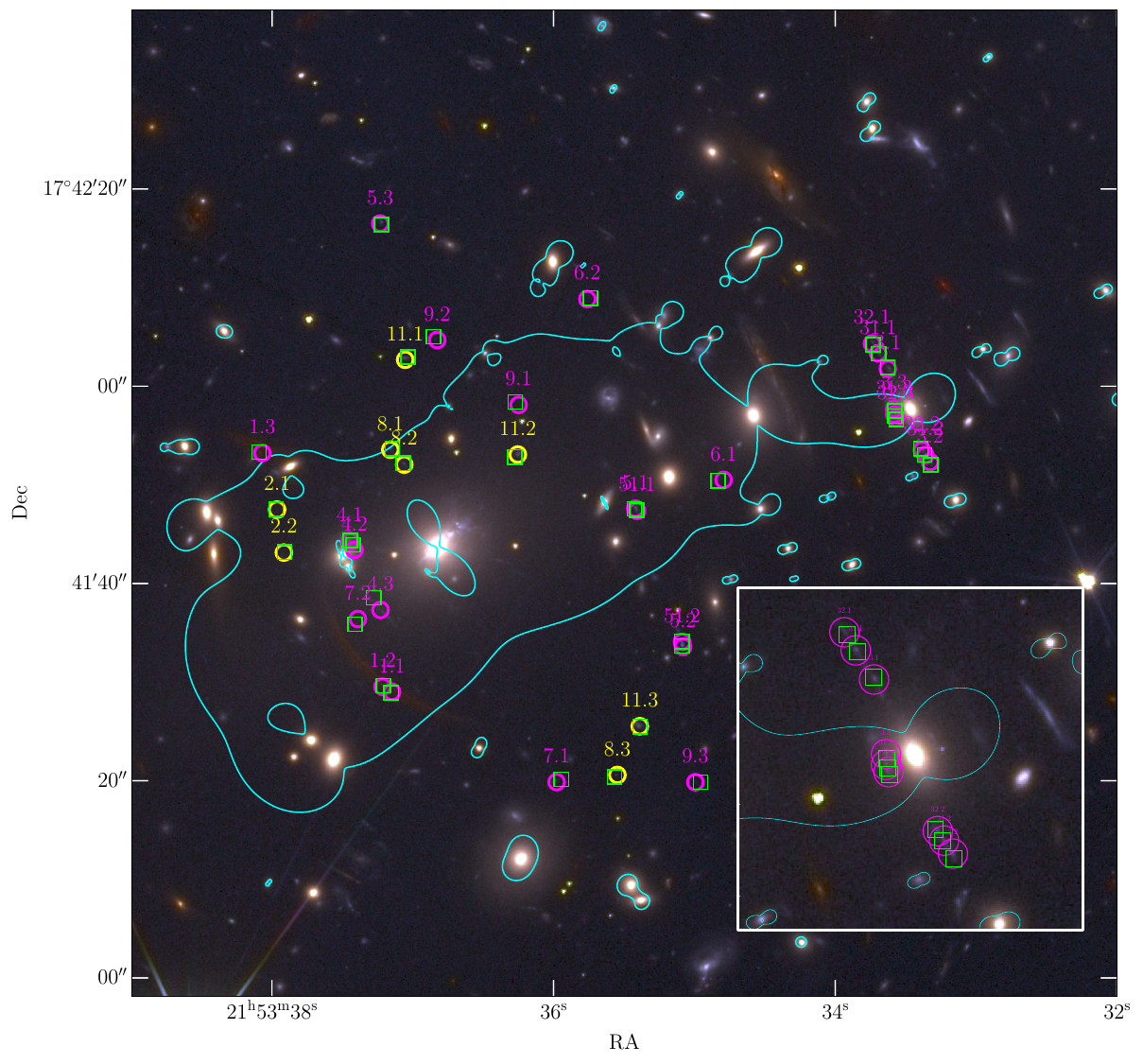}
    \caption{Colour-composite \textit{Euclid} image (red: $\HE$, green: $\JE$+$\HE$, blue: $\IE$) of the galaxy cluster A2390 with the critical lines from the reference model M9 evaluated for a source at redshift $z_\mathrm{s} = 4.048$ (families 3, 31, and 32) overlaid in cyan. The spectroscopically confirmed and photometric multiple images included in our analysis are also shown in magenta and yellow, respectively. Green boxes denote the predictions of the reference model. The bottom-right insert depicts a zoom-in of the area around G29.}
    \label{fig:cc}
\end{figure*}

Model M9 is characterised by a precision of $\Delta_\mathrm{RMS} = \ang{;;0.32}$ in reproducing the observed positions of the 35 multiple images used to build the lens models. In Fig. \ref{fig:cc}, the model-predicted positions of the multiple images are shown as green boxes, and the observed spectroscopically confirmed (photometric) images are displayed as magenta (yellow) circles. Figure \ref{fig:cc} also shows the critical curves evaluated at the redshift of families 3, 31, and 32, at $z = 4.048$, namely, the multiple-image systems lying around G29. The bottom-right insert in the figure shows a zoom-in of the area around G29. A natural extension of the model would be to improve the number of constraints by modelling the surface brightness distribution of the arc where the above-mentioned families are observed.

\subsection{Total mass distribution}
\begin{figure}
    \centering
    \includegraphics[width=1.0\linewidth]{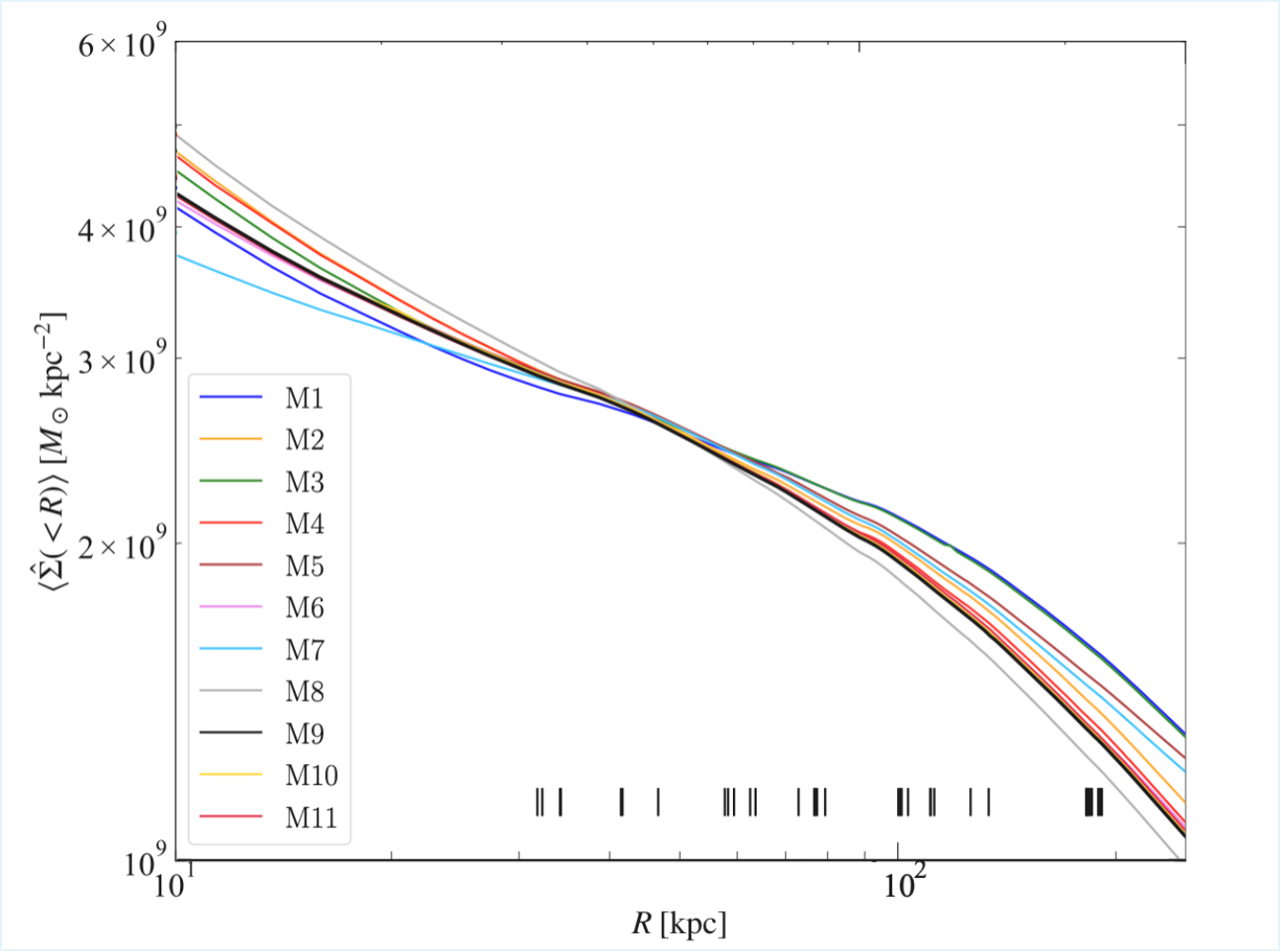}
    \caption{Average cumulative projected total surface mass density profile of A2390 from the reference (M9, in black) and the other models explored as a function of the projected distance from the BCG. The vertical black lines in the bottom part of the plot are the projected distances from the BCG of the 35 multiple images included in our models.}
    \label{fig:profili-di-massa}
\end{figure}

\begin{figure}
    \centering
    \includegraphics[width=1.0\linewidth]{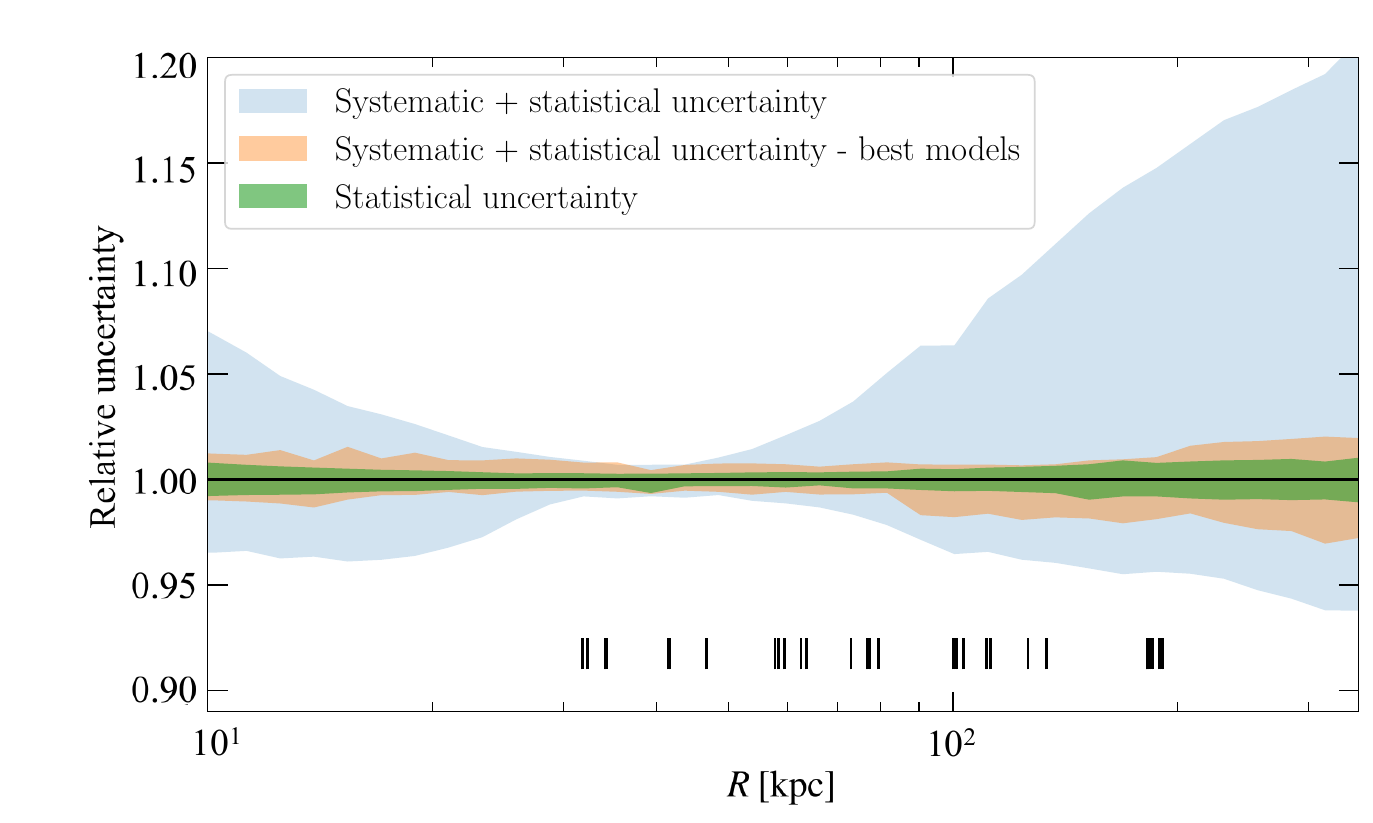}
    \caption{Relative statistical (green) and total systematic+statistical (blue and orange) uncertainty with respect to the cumulative total mass profile from the reference model. The total error is estimated by considering either all the 11 models explored (in blue) or the three best ones (in orange), according to the figures of merit adopted in this work, namely, M6, M9, and M11.}
    \label{fig:incertezze}
\end{figure}

We find that the cumulative total mass profiles from the 11 parametrisations are in excellent agreement with each other. This is expected since \cite{meneghetti2017} have showed that the total mass measured within the region where multiple images are observed is the quantity evaluated with better precision. In Fig.~\ref{fig:profili-di-massa}, for better visualisation, we show the average cumulative projected total surface mass density profiles of the cluster as a function of the distance from the BCG for the reference model of this work (M9) and for the other models explored. As emerges from Fig.~\ref{fig:profili-di-massa}, the parametrisation of model M9 leads to an isothermal fall-off at large radii. This is expected, as in regions distant from the core of the cluster, no multiple images are observed, and the predictions are thus an extrapolation of the NIE profile adopted. The observed density profile is inconsistent with the Navarro, Frenk, and White one \citep{navarro96, navarro97} expected for cluster haloes \citep{wang2020}, but this may be attributed to the extrapolation cited above. The predictions of our model are less robust when far from the SL region. We find a projected total mass value of $M(< 40 \; \mathrm{kpc}) = (1.40\pm0.01) \times 10^{13} \, M_\odot$ within the projected distance of the multiple images from the BCG at which we found the lowest uncertainty (see Fig. \ref{fig:incertezze}). The upper and lower limits represent the statistical uncertainty, evaluated as the 84th and 16th percentiles, respectively, estimated by generating 500 total mass profiles from 500 sets of parameters randomly extracted from the MCMC realisations of M9. For model M9, we also estimated the effective Einstein radius, $\theta_\mathrm{E}$, as
\begin{equation}
    \theta_\mathrm{E} = \sqrt{\frac{A_\mathrm{c}}{\pi}} \, ,
\end{equation}
with $A_\mathrm{c}$ as the area enclosed by the principal critical curve. At redshift $z = 4.048$, we find $\theta_\mathrm{E} = (24.6\pm 5.4)''$. We randomly extracted 100 points from the MCMC realisations, and for each, we estimated the corresponding $\theta_\mathrm{E}$. We quote the median and the uncertainty, evaluated as the semi-difference between the 84th and the 16th quantiles.
Thanks to the exploration of the other models, we were able to quantify the systematic uncertainty arising from our modelling choices. For each model, including M9, we generated 500 realisations of the total mass profile, similar to the procedure followed for M9 only. The result, displaying the relative impact of statistical and systematic uncertainties, is shown in Fig. \ref{fig:incertezze}, where the blue (green) band corresponds to the interval [16th, 84th] percentiles associated with the total statistical+systematic (statistical only) uncertainty. As is visible in the figure, the uncertainty budget on the cluster total mass is mainly dominated by systematic effects. The total systematic+statistical uncertainty (blue band) can be as large as $15\%$ in the outermost SL region (where the most distant multiple images are observed), with the statistical one (green band) counting up to $0.8\%$. This systematic uncertainty was estimated by taking into account all the models studied, with no weighing applied according to their RMS or evidence. Nevertheless, if we restrict the sample to the best three models (according to these figures of merit), namely, models M6, M9, and M11, the systematic effects drastically reduce to roughly $1.5\%$ (see the orange band in Fig. \ref{fig:incertezze}).

\subsection{Comparison with weak lensing}
\begin{figure}
    \centering
    \includegraphics[width=1.0\linewidth]{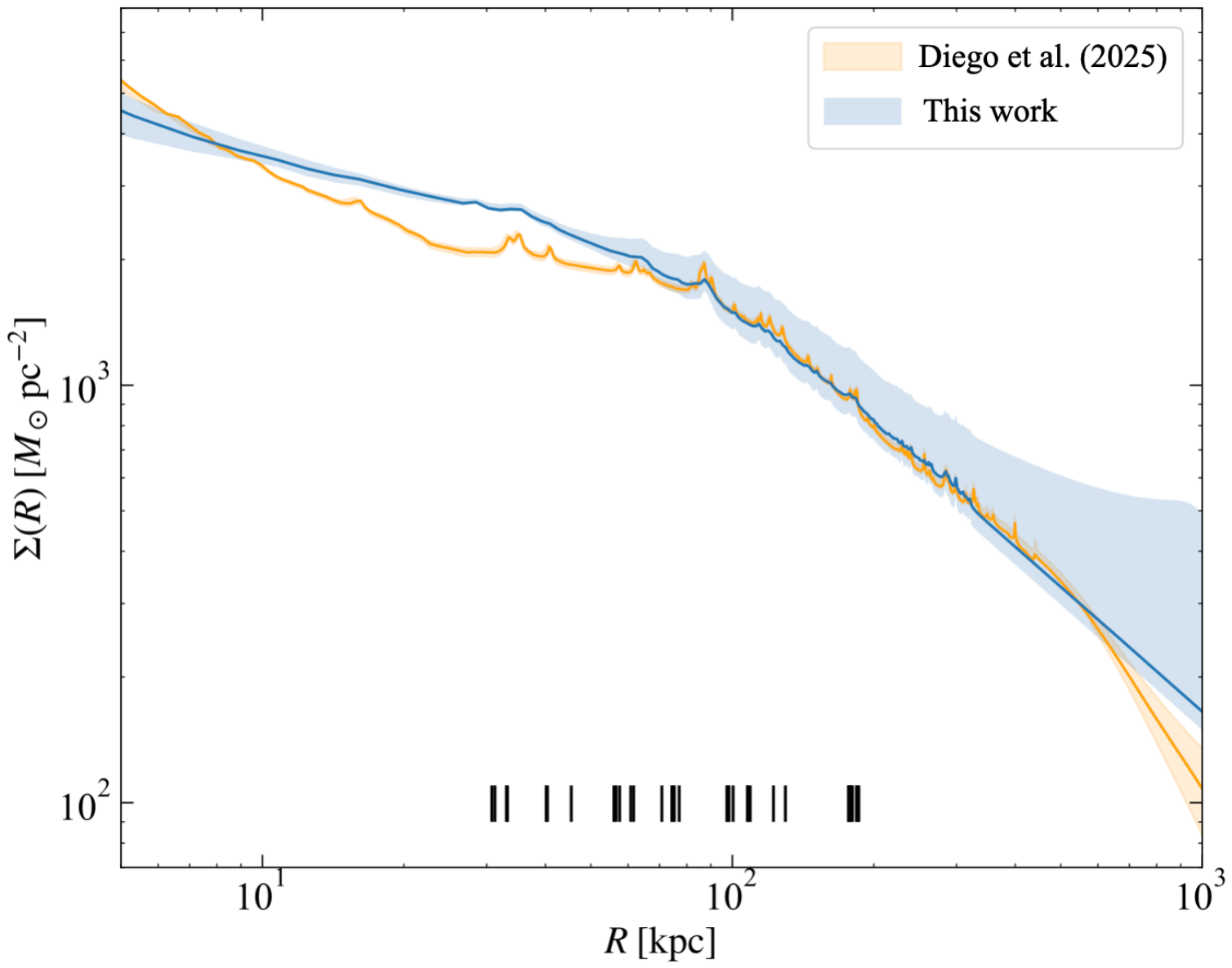}
    \caption{Surface mass density profile of A2390 as a function of the distance from the BCG. The blue (orange) solid line corresponds to this work (the lens model by Diego et al., in prep.). The light blue area quantifies the systematic uncertainty enclosing the minimum and maximum values of the mass profiles for all 11 models explored in this work. The light orange area is estimated as described in \cite{diego2390}. The vertical black lines in the bottom part of the plot are the projected distances from the BCG of the 35 multiple images included in our models.}
    \label{fig:comp}
\end{figure}

Next, we compared our work with the joint strong and weak lensing analysis first presented in \cite{atekero} and then developed in \cite{diego2390}, obtained with the free-form software \texttt{WSLAP+} \citep{diego2005, diego2007} based on the same \textit{Euclid} imaging. Differently from ours, the study relies on a free-form approach and exploits a different set of multiple images to reconstruct the cluster total mass distribution. In Fig. \ref{fig:comp}, we compare the surface mass density profile obtained in our work (blue curve) with the results from Diego et al. (in prep., orange curve). For a fair comparison, we restricted ourselves to the region where multiple images (represented with vertical black lines) are observed. In the outer regions of the cluster, where no SL features are present, the extrapolation of our reference model indeed becomes less robust. We found a nice agreement in the region between approximately 60 kpc and 500 kpc from the BCG. The discrepancy observed between the two total surface mass density profiles in the innermost (a few tens of kiloparsecs from the BCG) and outermost (a few hundreds of kiloparsecs) regions may be ascribed to the lack of SL features. In particular, in the innermost SL regime, we think that the disagreement can mainly be attributed to the different approach adopted to reconstruct the total mass distribution of the cluster. \cite{meneghetti2017} have found that parametric models are generally more accurate and precise in reconstructing the projected total mass density of a lens cluster.

\section{Conclusions}
\label{conclusions}

\noindent We have presented new high-precision strong lensing modelling of the \Euclid ERO galaxy cluster Abell 2390 ($z$ = 0.231). To build our model, we combined, for the first time, the Euclid VIS and NISP observations with spectroscopy from MUSE archival data, which we fully re-analysed here. We identified 35 multiple images from 13 background families, 25 of which are spectroscopically confirmed, spanning a redshift range from $z = 0.535$ to $z = 4.877$. We included in our model 65 secure cluster members and added 114 photometric members by studying their distribution in colour and magnitude. We were able to measure the stellar velocity dispersion for 22 cluster members, which allowed us to properly calibrate the sub-halo scaling relations used in the modelling and alleviate inherent degeneracies between the cluster- and the galaxy-scale mass components. We performed our analysis with the new software \texttt{Gravity.jl}, with which we explored 11 parametrisations of the total mass distribution of the galaxy cluster with an increasing level of complexity. 

The cluster is best described with a single large-scale smooth halo and an external shear term. Our reference model is characterised by a total RMS separation between the observed and model-predicted positions of the 35 multiple images of $\Delta_\mathrm{RMS} = \ang{;;0.32}$. We were able to reconstruct the total mass distribution of the galaxy cluster and estimate the systematic uncertainty arising from our modelling choices by taking advantage of all the models efficiently explored with \texttt{Gravity.jl}. On average, full optimisation of a mass model required approximately 2 hours to be obtained on a 64-core workstation, thus allowing us to test different total mass models in a very limited amount of time. 

Notably, \texttt{Gravity.jl} allows for fast and reliable total mass reconstruction of cluster lenses. Thus, it is an ideal  tool for SL analyses of large samples of galaxy clusters, such as the one that \Euclid will deliver. Indeed, based on the forecasts by \cite{boldrin2012, boldrin2016}, we expect that \Euclid will observe SL features in more than $6000$ galaxy clusters. These estimates are consistent with the number of SL clusters found in the Q1 data release \citep{b25}. At the same time, \texttt{Gravity.jl} will also speed up the analysis of clusters observed at a greater depth compared to \Euclid observations. As suggested by recent JWST observations, deep follow-up observations of clusters identified in the \Euclid surveys will likely reveal hundreds of families of multiple images. These observations, combined with spectroscopic follow-up, will enable detailed and robust total mass models with \texttt{Gravity.jl}. This study proves that with \texttt{Gravity.jl}, robust SL analyses can be achieved in a short amount of time.

\begin{acknowledgements}
This work has made use of data from the European Space Agency (ESA) mission Euclid (\href{https://www.cosmos.esa.int/euclid}{https://www.cosmos.esa.int/euclid}), processed by the Euclid Consortium (\href{https://www.euclid-ec.org/}{https://www.euclid-ec.org/}). This work has also made use of the Early Release Observation (ERO) data from the European Space Agency (ESA) mission Euclid, available at \href{https://euclid.esac.esa.int/dr/ero/}{https://euclid.esac.esa.int/dr/ero/}. This work is based on observations collected at the European Southern Observatory under ESO programme with ID 094.A-0115 (P.I.: Richard). GA acknowledges financial contributions from the agreements of the Euclid ESA mission – ASI/INAF 2018-23-HH.0 Phase D and ASI-INAF n. 2024-10-HH.0 - Phase E. This project has been partially funded under the National Recovery and Resilience Plan (NRRP), funded by the European Union – NextGenerationEU; Project title “GRAVITY”, project code PNRR-BAC24MLOMB-01, CUP C53C22000350006.
\end{acknowledgements}

\bibliography{biblio}

\begin{appendix}

\section{Total mass parametrisations}
\label{appA}
In Table \ref{table:lens_models} we present a brief summary of the lens models explored in this work, M1 to M11. For each of them, a brief description of the parametrisation is given, as well as the number of free parameters, the degrees of freedom, and the two figures of merit used to quantify their goodness adopted in our analysis, the RMS $\Delta_\mathrm{RMS}$ and the evidence of Bayes' theorem.

\begin{table}[htb!]  
    \caption{Description of the lens models explored in this work.}
	\tiny
	\centering    
	\begin{tabular}{c c c c c l}
       \noalign{\vskip 2pt}
	   \textbf{\normalsize Model} & \boldmath{\normalsize $N_\mathrm{par}$} & \textbf{\normalsize $N_\mathrm{dof}$} & \boldmath{\normalsize $\Delta_\mathrm{RMS}\,\mathrm{[\arcsec]}$} &
	   \boldmath{\normalsize $- \ln E$} &
	   \textbf{\normalsize Description} \\
	   \hline \noalign{\vskip 2pt}
	   \textbf{M1} & 11 & 33 & 0.52 & 592 & One cluster-scale DM halo, 179 sub-halo components \cr
	   \noalign{\vskip 4pt}
	   $\textbf{M2}$ & 13 & 31 & 0.40 & 476 & One cluster-scale DM halo, 179 sub-halo components, an external shear-like term \cr
	  \noalign{\vskip 4pt} 
        $\textbf{M3}$ & 13 & 31 & 0.53 & 651 & One cluster-scale DM halo, 179 sub-halo components, with the BCG excluded from the scaling \\
        & & & & & relations \cr
        \noalign{\vskip 4pt}
        $\textbf{M4}$ & 15 & 29 & 0.39 & 479 & One cluster-scale DM halo, 179 sub-halo components, with the BCG excluded from the scaling\\
        & & & & & relations, an external shear-like term \cr
        \noalign{\vskip 4pt}    
        \textbf{M5} & 17 & 27 & 0.41 & 445 & Two cluster-scale DM haloes, 179 sub-halo components \cr
	\noalign{\vskip 4pt}   
	   $\textbf{M6}$ & 19 & 25 & 0.31 & 379 & Two cluster-scale DM haloes, 179 sub-halo components, an external shear-like term \cr
	 \noalign{\vskip 4pt}  
        $\textbf{M7}$ & 19 & 25 & 0.41 & 461 & Two cluster-scale DM haloes, 179 sub-halo components, with the BCG excluded from the scaling\\
        & & & & & relations \cr
       \noalign{\vskip 4pt} 
        $\textbf{M8}$ & 21 & 23 & 0.34 & 393 & Two cluster-scale DM haloes, 179 sub-halo components, with the BCG excluded from the scaling\\
        & & & & & relations, an external shear-like term \cr
        \noalign{\vskip 4pt}
        $\textbf{M9}$ & 15 & 29 & 0.32 & 397 & One cluster-scale DM halo, 179 sub-halo components, an external shear-like term, with galaxy G29 \\
        & & & & & excluded from the scaling relations (spherical dPIE density profile)\cr
       \noalign{\vskip 4pt} 
        $\textbf{M10}$ & 17 & 27 & 0.32 & 399 & One cluster-scale DM halo, 179 sub-halo components, an external shear-like term, with galaxy G29 \\
        & & & & & excluded from the scaling relations (elliptical dPIE density profile)\cr
        \noalign{\vskip 4pt}
        $\textbf{M11}$ & 19 & 25 & 0.27 & 363 & Three cluster-scale DM haloes, 179 sub-halo components, an external shear-like term \cr
    
	\end{tabular}
    \tablefoot{$N_\mathrm{par}$ and $N_\mathrm{dof}$ are the number of model free-parameters and the degrees-of-freedom, respectively. $\Delta_\mathrm{RMS}$ is the RMS displacement between the positions of observed and model-predicted multiple images. $\ln E$ is the natural logarithm of the evidence of Bayes' theorem for each model. In the last column, we briefly summarise the total mass parametrisation of the lens model.}
	\smallskip
	\label{table:lens_models} 

\end{table}

\begin{figure*}[hbt!]
    \centering
    \includegraphics[width=1.0\linewidth]{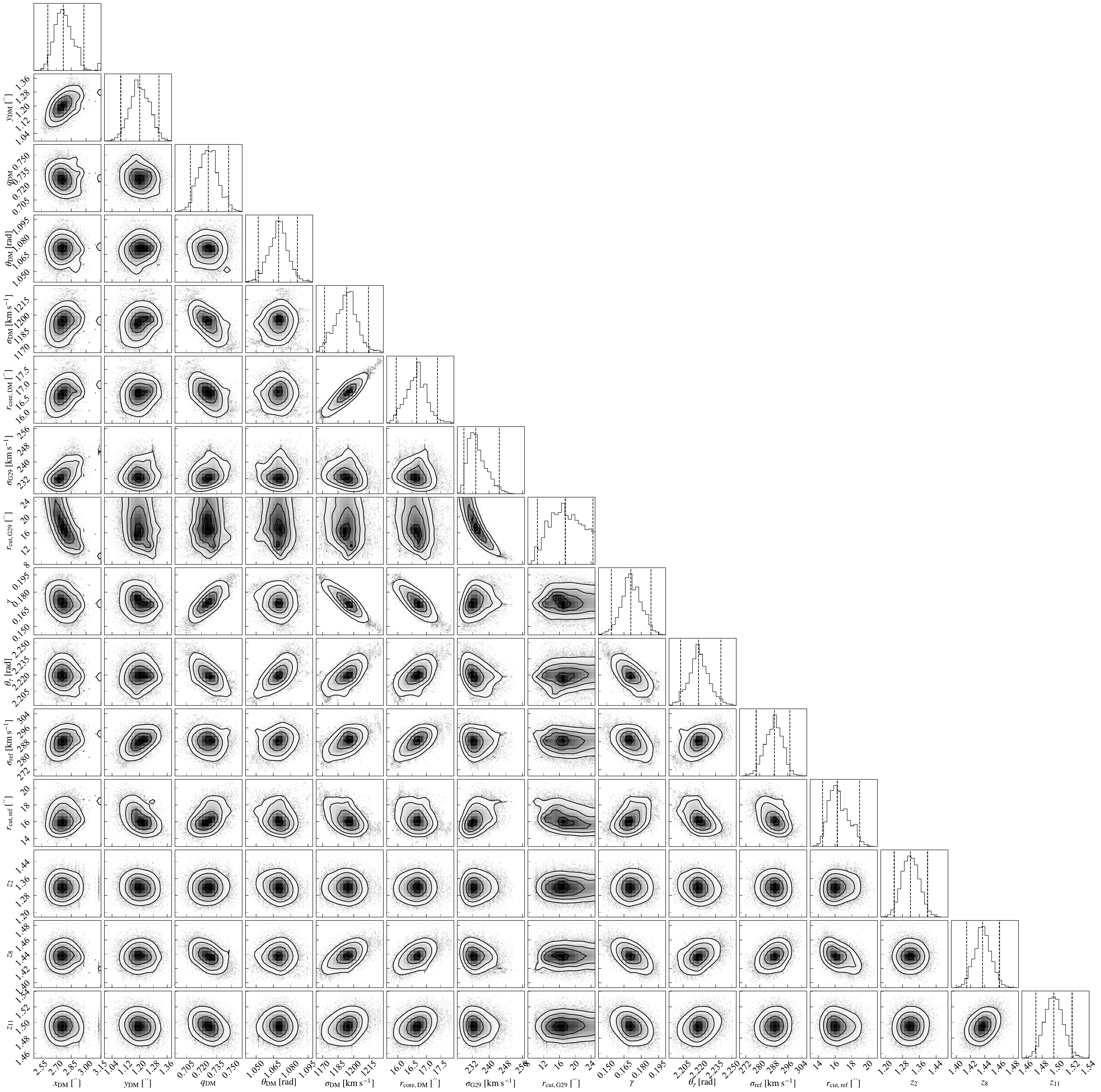}
    \caption{Marginalised posterior distributions of the parameters of the reference model of this work, M9. The 16th, 50th and
84th percentiles of the marginalised distributions are shown as vertical dashed lines.}
    \label{fig:cornerplotM9}
\end{figure*}

\FloatBarrier
\section{Cluster members catalogue}
In this Appendix, we present the catalogue of the cluster members included in our total mass parametrisations. Table \ref{tab:specmems1} lists the 65 spectroscopically confirmed cluster members, identified by their ID, position, spectroscopic redshift, and $\HE$ magnitude. Table \ref{tab:photmems1} lists the 114 photometric cluster members, for which we report their position and $\HE$ magnitude.

\begin{table}[hbt!]
\caption{ID, coordinates, spectroscopic redshift, and $\HE$ magnitude of the spectroscopic cluster members included in the models explored in this work.}
\label{tab:specmems1}
\centering
\begin{tabular}{ccccc} 
  ID & RA & Dec & $z_\mathrm{spec}$ & $\HE$\\ 
& [$\mathrm{deg}$] & [$\mathrm{deg}$] &  & \\
\noalign{\vskip 2pt}
 \hline
 \noalign{\vskip 2pt}
1 & 328.403418 & 17.695474 & 0.2301 & 15.37 \\
129 & 328.400980 & 17.686700 & 0.2465 & 16.75 \\
29 & 328.389480 & 17.699362 & 0.2243 & 16.84 \\
132 & 328.406506 & 17.689503 & 0.2294 & 16.90 \\
23 & 328.394113 & 17.699178 & 0.2301 & 17.10 \\
6 & 328.390237 & 17.690741 & 0.2284 & 17.16 \\
13 & 328.396608 & 17.697441 & 0.2317 & 17.22 \\
135 & 328.380951 & 17.692656 & 0.2459 & 17.34 \\
139 & 328.428199 & 17.698142 & 0.2238 & 17.57 \\
65 & 328.394020 & 17.703793 & 0.2263 & 17.61 \\
123 & 328.419564 & 17.680547 & 0.2313 & 17.67 \\
117 & 328.403859 & 17.676102 & 0.2275 & 17.69 \\
145 & 328.424179 & 17.704111 & 0.2433 & 17.72 \\
5 & 328.406259 & 17.695223 & 0.2280 & 17.73 \\
138 & 328.410274 & 17.696451 & 0.2311 & 17.78 \\
128 & 328.397792 & 17.685917 & 0.2278 & 17.84 \\
58 & 328.400041 & 17.703490 & 0.2317 & 17.86 \\
118 & 328.416810 & 17.676787 & 0.2404 & 17.89 \\
134 & 328.407167 & 17.690000 & 0.2335 & 17.95 \\
555 & 328.426530 & 17.690097 & 0.2263 & 18.12 \\
144 & 328.420311 & 17.703075 & 0.2226 & 18.15 \\
142 & 328.383097 & 17.699696 & 0.2372 & 18.27 \\
456 & 328.407807 & 17.697758 & 0.2346 & 18.29 \\
116 & 328.414478 & 17.673639 & 0.2353 & 18.29 \\
140 & 328.410913 & 17.698304 & 0.2290 & 18.62 \\
121 & 328.422178 & 17.679424 & 0.2155 & 18.63 \\
549 & 328.390609 & 17.707208 & 0.2320 & 18.77 \\
37 & 328.388124 & 17.696791 & 0.2255 & 18.80 \\
33 & 328.398502 & 17.696725 & 0.2346 & 18.81 \\
481 & 328.391098 & 17.714808 & 0.2321 & 18.84 \\
21 & 328.406071 & 17.694906 & 0.2275 & 18.89 \\
143 & 328.409748 & 17.701550 & 0.2275 & 18.89 \\
54 & 328.386584 & 17.700838 & 0.2348 & 18.97 \\
454 & 328.401332 & 17.673742 & 0.2391 & 19.00 \\
452 & 328.383745 & 17.702682 & 0.2319 & 19.04 \\
34 & 328.393902 & 17.696538 & 0.2278 & 19.08 \\
48 & 328.403036 & 17.698513 & 0.2260 & 19.09 \\
126 & 328.392682 & 17.684349 & 0.2312 & 19.13 \\
22 & 328.398174 & 17.695619 & 0.2316 & 19.22 \\
501 & 328.415601 & 17.707220 & 0.2339 & 19.27 \\
41 & 328.396935 & 17.698578 & 0.2201 & 19.30 \\
27 & 328.391192 & 17.694975 & 0.2299 & 19.34 \\
15 & 328.407495 & 17.693431 & 0.2422 & 19.45 \\
8 & 328.402209 & 17.689812 & 0.2262 & 19.58 \\
59 & 328.396907 & 17.701705 & 0.2411 & 19.66 \\
14 & 328.407482 & 17.692931 & 0.2446 & 19.79 \\
32 & 328.393082 & 17.695434 & 0.2268 & 19.81 \\
61 & 328.387322 & 17.701042 & 0.2246 & 19.97 \\
49 & 328.398382 & 17.699659 & 0.2284 & 20.08 \\
46 & 328.389401 & 17.697469 & 0.2418 & 20.26 \\
28 & 328.404712 & 17.695246 & 0.2313 & 20.27 \\
63 & 328.392174 & 17.700680 & 0.2303 & 20.50 \\
25 & 328.394782 & 17.694560 & 0.2294 & 20.52 \\
62 & 328.402033 & 17.700927 & 0.2235 & 20.56 \\
39 & 328.391895 & 17.696872 & 0.2208 & 20.59 \\
\end{tabular}
\end{table}

\begin{table}[!ht]
\label{tab:specmems2}
\centering
\begin{tabular}{ccccc} 

  ID & RA & Dec & $z_\mathrm{spec}$ & $\HE$\\ 
& [$\mathrm{deg}$] & [$\mathrm{deg}$] &  & \\
\noalign{\vskip 2pt}
 \hline
 \noalign{\vskip 2pt}
64 & 328.395390 & 17.701352 & 0.2319 & 20.60 \\
35 & 328.401145 & 17.695504 & 0.2318 & 20.62 \\
66 & 328.396756 & 17.701983 & 0.2241 & 20.73 \\
18 & 328.403713 & 17.694764 & 0.2306 & 20.86 \\
31 & 328.389208 & 17.695613 & 0.2334 & 20.92 \\
17 & 328.390124 & 17.693444 & 0.2404 & 21.27 \\
57 & 328.403111 & 17.699426 & 0.2257 & 21.59 \\
12 & 328.396655 & 17.692354 & 0.2373 & 22.05 \\
69 & 328.394812 & 17.703081 & 0.2324 & 22.15 \\
75 & 328.400852 & 17.705477 & 0.2375 & 22.31 \\

\end{tabular}
\end{table}

\begin{table}[!ht]
\caption{Coordinates and $\HE$ magnitude of the photometric cluster members included in the models explored in this work.}
\label{tab:photmems1}
\centering
\begin{tabular}{ccc} 

  RA & Dec & $\HE$ \\
  {[$\mathrm{deg}$]} & [$\mathrm{deg}$] & \\
  \noalign{\vskip 2pt}
 \hline
 \noalign{\vskip 2pt}
328.395373 & 17.679027 & 18.31 \\
328.419201 & 17.684186 & 18.42 \\
328.409937 & 17.696223 & 18.53 \\
328.397427 & 17.685539 & 18.56 \\
328.422078 & 17.691723 & 18.60 \\
328.425706 & 17.690011 & 18.65 \\
328.379492 & 17.690138 & 18.78 \\
328.390766 & 17.707999 & 18.87 \\
328.388423 & 17.712424 & 19.23 \\
328.387095 & 17.712811 & 19.24 \\
328.419327 & 17.708467 & 19.35 \\
328.420524 & 17.707458 & 19.41 \\
328.390916 & 17.679384 & 19.48 \\
328.398056 & 17.673459 & 19.48 \\
328.417843 & 17.679514 & 19.63 \\
328.423359 & 17.708350 & 19.66 \\
328.383745 & 17.681237 & 19.74 \\
328.421229 & 17.679324 & 20.08 \\
328.398580 & 17.710146 & 20.11 \\
328.407470 & 17.708739 & 20.18 \\
328.387165 & 17.709260 & 20.31 \\
328.388858 & 17.680391 & 20.32 \\
328.411416 & 17.698291 & 20.39 \\
328.398244 & 17.708383 & 20.52 \\
328.386466 & 17.688472 & 20.68 \\
328.397987 & 17.700801 & 20.75 \\
328.414557 & 17.690027 & 20.83 \\
328.392893 & 17.694584 & 21.04 \\
328.388050 & 17.707786 & 21.04 \\
328.395030 & 17.692832 & 21.04 \\
328.398450 & 17.697732 & 21.06 \\
328.378788 & 17.694668 & 21.10 \\
328.387660 & 17.708136 & 21.13 \\
328.406028 & 17.678437 & 21.18 \\
328.396279 & 17.705389 & 21.20 \\
328.395716 & 17.708390 & 21.27 \\
328.400771 & 17.709475 & 21.30 \\
328.412377 & 17.674895 & 21.34 \\
328.395140 & 17.705197 & 21.36 \\
328.421231 & 17.697061 & 21.37 \\
328.405221 & 17.712687 & 21.40 \\
328.400367 & 17.702921 & 21.61 \\
328.405150 & 17.687042 & 21.62 \\
328.412625 & 17.702028 & 21.63 \\
328.408438 & 17.686019 & 21.66 \\
328.425366 & 17.696199 & 21.67 \\
328.412807 & 17.711701 & 21.68 \\
328.411456 & 17.676374 & 21.71 \\
328.384534 & 17.710268 & 21.73 \\
328.393620 & 17.718180 & 21.75 \\
328.397974 & 17.684747 & 21.76 \\
328.418133 & 17.683454 & 21.80 \\
328.411837 & 17.709412 & 21.86 \\
328.417213 & 17.706249 & 21.91 \\
328.384429 & 17.681682 & 21.91 \\
328.416339 & 17.683650 & 21.93 \\

\end{tabular}
\end{table}

\begin{table}[!ht]
\label{tab:photmems2}
\centering
\begin{tabular}{ccc} 
  RA & Dec & $\HE$ \\ 
{[$\mathrm{deg}$]} & [$\mathrm{deg}$] & \\
\noalign{\vskip 2pt}
 \hline
 \noalign{\vskip 2pt}
328.379613 & 17.692168 & 21.96 \\
328.412754 & 17.702376 & 22.01 \\
328.387582 & 17.686001 & 22.01 \\
328.383801 & 17.698588 & 22.02 \\
328.417443 & 17.690444 & 22.03 \\
328.400688 & 17.674176 & 22.06 \\
328.401678 & 17.679097 & 22.08 \\
328.397523 & 17.690469 & 22.19 \\
328.414195 & 17.689926 & 22.19 \\
328.420850 & 17.681010 & 22.20 \\
328.390874 & 17.702680 & 22.20 \\
328.426985 & 17.698382 & 22.23 \\
328.392301 & 17.678618 & 22.23 \\
328.390167 & 17.692743 & 22.25 \\
328.417624 & 17.675773 & 22.28 \\
328.404124 & 17.688855 & 22.31 \\
328.386274 & 17.701900 & 22.31 \\
328.390865 & 17.682477 & 22.33 \\
328.395132 & 17.679975 & 22.34 \\
328.390577 & 17.685685 & 22.34 \\
328.394710 & 17.674228 & 22.35 \\
328.382411 & 17.703971 & 22.35 \\
328.402158 & 17.709105 & 22.36 \\
328.414735 & 17.697317 & 22.37 \\
328.404436 & 17.697823 & 22.42 \\
328.399016 & 17.709886 & 22.43 \\
328.417525 & 17.695840 & 22.44 \\
328.399131 & 17.703406 & 22.46 \\
328.411468 & 17.687562 & 22.48 \\
328.400644 & 17.677260 & 22.48 \\
328.394823 & 17.692407 & 22.52 \\
328.422040 & 17.686636 & 22.56 \\
328.399841 & 17.708115 & 22.57 \\
328.379403 & 17.692802 & 22.57 \\
328.424205 & 17.703253 & 22.59 \\
328.382970 & 17.698536 & 22.61 \\
328.384634 & 17.707335 & 22.65 \\
328.415015 & 17.687250 & 22.65 \\
328.392229 & 17.681096 & 22.66 \\
328.400346 & 17.708374 & 22.69 \\
328.398442 & 17.704258 & 22.71 \\
328.380145 & 17.703602 & 22.72 \\
328.391982 & 17.717290 & 22.73 \\
328.418263 & 17.684292 & 22.74 \\
328.415171 & 17.696093 & 22.75 \\
328.392433 & 17.693603 & 22.78 \\
328.424831 & 17.701809 & 22.80 \\
328.400695 & 17.678667 & 22.81 \\
328.407131 & 17.699966 & 22.81 \\
328.388916 & 17.687924 & 22.81 \\
328.381149 & 17.693650 & 22.85 \\
328.388185 & 17.715269 & 22.91 \\
328.398344 & 17.685101 & 22.93 \\
328.388076 & 17.678557 & 22.93 \\
328.382448 & 17.707600 & 22.94 \\
328.411479 & 17.694391 & 22.95 \\
328.385941 & 17.709318 & 22.97 \\
328.384893 & 17.688864 & 22.97 \\

\end{tabular}
\end{table}

\FloatBarrier
\section{Measured velocity dispersion catalogue}
In this Appendix, we present the final MUSE velocity dispersion catalogue, containing velocity dispersion measurements for 22 cluster members, as discussed in Sect. \ref{cluster-mems}.

\begin{table}[hbt!]
\caption{Catalogue of measured velocity dispersion values.}             
\label{tab:vdispcat}      
\centering          
\begin{tabular}{c c c c c c c }     
              
ID & RA & Dec & $z_\mathrm{spec}$ & $\sigma_\mathrm{gal}$ & $\delta \sigma_\mathrm{gal}$ & $\mathrm{\langle S/N \rangle}$ \\  
& [deg] & [deg] &  & [$\kms$] & [$\kms$] & \\ 
\noalign{\vskip 2pt}
\hline       
\noalign{\vskip 2pt}
1 & 328.403418 & 17.695474 & 0.2301 & 276.8 & 13.8\phantom{0} & 81.1 \\ 
3 & 328.395241 & 17.693922 & 0.5212 & 114.1 & 4.4 & 28.0 \\ 
5* & 328.406259 & 17.695223 & 0.2280 & 183.4 & 3.6 & 45.2 \\ 
8 & 328.402209 & 17.689812 & 0.2262 & 197.8 & 10.0\phantom{0} & 19.2 \\ 
13 & 328.396608 & 17.697441 & 0.2317 & 233.3 & 2.6 & 77.7 \\ 
15 & 328.407495 & 17.693431 & 0.2422 & 135.2 & 5.3 & 24.3 \\ 
21* & 328.406071 & 17.694906 & 0.2275 & 181.7 & 4.9 & 32.6 \\ 
22 & 328.398174 & 17.695619 & 0.2316 & \phantom{0}96.0 & 4.0 & 23.0 \\ 
23 & 328.394113 & 17.699178 & 0.2301 & 195.5 & 2.2 & 77.5 \\ 
27 & 328.391192 & 17.694975 & 0.2299 & 136.7 & 5.3 & 24.0 \\ 
28* & 328.404712 & 17.695246 & 0.2313 & 174.5 & 8.3 & 18.8 \\ 
29 & 328.389480 & 17.699362 & 0.2243 & 237.3 & 2.7 & 81.0 \\ 
32 & 328.393082 & 17.695434 & 0.2268 & \phantom{0}95.7 & 5.1 & 17.6 \\ 
33 & 328.398502 & 17.696725 & 0.2346 & 126.8 & 3.4 & 32.7 \\ 
34 & 328.393902 & 17.696538 & 0.2278 & \phantom{0}93.7 & 3.9 & 24.4 \\ 
35 & 328.401145 & 17.695504 & 0.2318 & 147.4 & 12.2\phantom{0} & 12.7 \\ 
41 & 328.396935 & 17.698578 & 0.2201 & \phantom{0}90.3 & 5.0 & 21.8 \\ 
48 & 328.403036 & 17.698513 & 0.2260 & 215.0 & 7.1 & 33.3 \\ 
58 & 328.400041 & 17.703490 & 0.2317 & 230.0 & 3.4 & 56.3 \\ 
59 & 328.396907 & 17.701705 & 0.2411 & \phantom{0}81.4 & 6.3 & 18.6 \\ 
61 & 328.387322 & 17.701042 & 0.2246 & \phantom{0}99.4 & 6.1 & 17.2 \\ 
65 & 328.394020 & 17.703793 & 0.2263 & 195.1 & 3.2 & 49.5 \\ 
136 & 328.390237 & 17.690741 & 0.2279 & 159.1 & 11.9\phantom{0} & 10.1 \\ 
                 
\end{tabular}
\tablefoot{We identify the galaxies included in this catalogue with their ID. We report the spectroscopic redshift of the galaxy $z_\mathrm{spec}$ (fourth column), its measured velocity dispersion value with $\sigma_\mathrm{gal}$ (fifth column), its uncertainty with $\delta \sigma_\mathrm{gal}$ (sixth column), and the spectral $\mathrm{\langle S/N \rangle}$ (seventh column). We mark the members potentially affected by light blending, for which we used a smaller aperture ($0.6''$ in radius, see Sect. \ref{cluster-mems}) for the spectral extraction, with *.}
\label{LastPage}
\end{table}

\end{appendix}

\end{document}